\documentclass[12pt,hidelinks]{article}
\usepackage{geometry}
\RequirePackage{amsthm,amsmath,amsfonts,amssymb, amsthm}
\RequirePackage[authoryear]{natbib}
\usepackage{amsmath}

\usepackage{lmodern}
\usepackage[T1]{fontenc}
\usepackage{bm}
\usepackage{natbib}
\usepackage{tikz}
\usetikzlibrary{fit, positioning}
\newcommand\sdots{\hbox to 0.5em{.\hss.\hss.}}
\usepackage[labelformat=simple]{subcaption}

\usepackage{bbm}
\usepackage{color,xcolor,soul}
\usepackage{graphicx}
\usepackage{diagbox}
\usepackage{pdflscape}
\usepackage{rotating}
\usepackage{varwidth}

\usepackage[plain,noend]{algorithm2e}
\usepackage{caption}
\usepackage{booktabs}
\usepackage{multirow}
\usepackage{hyperref}
\usepackage{authblk}

\usepackage[capitalize]{cleveref}

\usepackage{mathtools} 

\title{Bayesian Poisson Regression and Tensor Train Decomposition Model for Learning Mortality Pattern Changes during COVID-19 Pandemic}
	
\author[1]{Wei Zhang\footnote{{wei.zhang@usi.ch}}}
\author[1]{Antonietta Mira\footnote{{antonietta.mira@usi.ch}}}
\author[1]{Ernst C. Wit\footnote{{ernst.jan.camiel.wit@usi.ch}}}

\affil[1]{Università della Svizzera italiana}

\begin{document}
\maketitle

\begin{abstract}
COVID-19 has led to excess deaths around the world, however it remains unclear how the mortality of other causes of death has changed during the pandemic. Aiming at understanding the wider impact of COVID-19 on other death causes, we study Italian data set that consists of monthly mortality counts of different causes from January 2015 to December 2020. Due to the high dimensional nature of the data, we develop a model which combines conventional Poisson regression with tensor train decomposition to explore the lower dimensional residual structure of the data. We take a Bayesian approach, impose priors on model parameters. Posterior inference is performed using an efficient Metropolis-Hastings within Gibbs algorithm. The validity of our approach is tested in simulation studies. Our method not only identifies differential effects of interventions on cause specific mortality rates through the Poisson regression component, but also offers informative interpretations of the relationship between COVID-19 and other causes of death as well as latent classes that underline demographic characteristics, temporal patterns and causes of death respectively.
\end{abstract}


\section{Introduction}
\label{Introduction}

Following the outbreak, COVID-19 has led to far-reaching consequences on various aspects of the world \citep{gormsen2020coronavirus, cheval2020observed, sarkodie2021global, kuzemko2020covid, bol2021effect}. Focusing on its impacts on health and health systems, extensive studies have investigated topics such as health inequality as a result of racial and social-economic statues \citep{bambra2020covid, abedi2021racial}, adaptation of health care system in terms of testing, contact tracing and vaccination campaign \citep{kretzschmar2020impact, peretti2020future}. Excess mortality due to the pandemic is also under scrutiny as it generates the overall picture of the impacts the pandemic has on the human health through various channels like government lockdown interventions, disruptions to non-COVID care and son on. The mortality pattern shift compared to pre-COVID era depends on the potential joint effects of all these factors \citep{karlinsky2021tracking, wang2022estimating, msemburi2023estimates}. Even though excess mortality is sufficient to grasp the general view, it is also of great importance to examine cause specific mortality changes in face of the pandemic so that strategies to mitigate similar impacts in the future can be more targeted. For instance, the pandemic may have indirectly led to increases in causes of death including heart disease, diabetes and Alzheimer disease as observed by \citet{shiels2022leading}. As for non-natural causes of death, \citet{dmetrichuk2022retrospective} found out that accidental drug-related fatalities increased substantially while homicide or suicide rates only moderately, nor did motor vehicle collision fatality rates greatly decrease during all stages of the lockdown in Ontario. However, evidences also suggest that suicide rates increased during the pandemic \citep{mitchell2021state, pell2020coronial} whereas deaths related to traffic accidents decreased significantly according to \citet{calderon2021impact} and \citet{sutherland2020vehicle}. However, it is usually challenging to collect cause specific mortality data based on death certificates in a consistent manner \citep{gill2020importance, gundlapalli2021death}. We analyze the Italian monthly death counts from 2015 to 2020 categorized according to the International Classification of Diseases 10th Revision (ICD-10), see Section \ref{Real Data Application} for more data description.

When the count data are assumed to follow Poisson distributions, the Poisson regression model is a good starting point \citep{frome1983analysis}. In practice, we can exploit other properties of the data and develop more sophisticated modeling tools in addition to the Poisson regression so that we learn more from the data. This is particularly important when the dimension of the observations is large or when we are not able to observe, collect all relevant covariates or when we suspect more complicated relationships between covariates and the outcome variable. The Italian mortality data can be rearrange as a multi-way array or tensor which facilitates us to subtract extra information hidden in the data thanks to its well studied theoretical properties, in fact, tensors have been shown to be a powerful tool in many disciplines such as political sciences, biology, economics and so on \citep{hoff2015multilinear, zhou2015bayesian, cai2022generalized}, therefore we utilize those properties, combine the Poisson regression with the tensor perspective in this applied work. Our primary interests lie in understanding the effects of covariates, especially government lockdown policies during the pandemic, on the mortality rates of various causes of death through the Poisson regression specification as well as uncovering further information in the data by inferring latent spaces via the tensor construction. Inferences are made in a Bayesian framework where we impose trivial priors on model parameters and employ a Metropolis within Gibbs sampler to draw posterior samples.

The rest of the paper is organized as follows. In Section \ref{Model}, we formulate the model and elucidate how to obtain dimension reduction via a tensor train decomposition. In Section \ref{Posterior}, we describe the prior specification and the Markov chain Monte Carlo (MCMC) algorithm for posterior inferences. Results of the simulation studies as well as the real data application are shown in Section \ref{Simulation Studies} and Section \ref{Real Data Application}, respectively.  Finally, Section \ref{Summary and Future Work} provides some concluding remarks and future work.

\section{Bayesian Poisson Regression and Tensor Train Decomposition model for count data}
\label{Model}

When high-dimensional data can be organized as tensors, to achieve dimension reduction and exploit inherent structure embedded in the data, researchers have developed numerous decomposition techniques. In this paper, we introduce the tensor train decomposition which has both theoretical and practical advantages \citep{oseledets2011tensor, cichocki2016tensor}. In general, an $M$-dimensional tensor $\mathcal{A}$ of size $Q_1\times Q_2\times \dots \times Q_M$ is said to admit a train decomposition if entries $a_{q_1,q_2,\dots,q_M}$ of $\mathcal{A}$ can be expressed as the sum of $R_1R_2\cdots R_{M-1}$ terms such that
\begin{equation*}
    a_{q_1,q_2,\dots,q_M} = \sum_{r_1=1}^{R_1} \sum_{r_2=1}^{R_2}\cdots \sum_{r_{M-1}=1}^{R_{M-1}}g^{(1)}_{q_1,r_1}g^{(2)}_{q_2,r_1,r_2}\cdots g^{(M)}_{q_M,r_{M-1}}.
\end{equation*}
We call $g^{(1)}_{\cdot,\cdot}, g^{(2)}_{\cdot,\cdot,\cdot}, \cdots, g^{(M)}_{\cdot,\cdot}$ tensor train cores and $R_1,R_2,\dots,R_{M-1}$ the tensor train ranks. The order of dimensions in the tensor matters as the decomposition is performed sequentially from the first dimension $g^{(1)}_{q_1,r_1}$ to the last dimension $g^{(M)}_{q_M,r_{M-1}}$ by construction, and tensor trains cores of a certain dimension always depend on the cores of its previous dimension. Therefore it is important to arrange the data in such a tensor structure that the train decomposition is meaningful. See Section \ref{Real Data Application} when we describe and analyze the Italian monthly cause specific mortality data. The tensor train decomposition has the theoretical advantages that it encompasses any specific tensor decomposition such as the canonical polyadic (CP) decomposition and the Tucker decomposition, but remains one of the most stable and simple approaches to summarize high-dimensional data by a limited number of latent variables, hence enabling straightforward interpretation of the results in application. We value these merits of the tensor train decomposition and employ it in our proposed model described as follows.

Suppose that we observe count data that can be arranged as a three-way discrete-valued tensor $Y_{i,t,k}$ of dimension $N\times T\times K$ and $i=1,\dots,N, t=1,\dots,T, k=1,\dots,K$. Additionally, we have information on covariates $\mathbf{x}_{i,t,k}\in \mathbb{R}^P$ and offsets $u_{i,t,k}$. Classical Poisson regression model assumes that
\begin{equation}
    Y_{i,t,k} \sim \text{Pois}\left(u_{i,t,k}\exp\left(\mathbf{x}_{i,t,k}\cdot\boldsymbol{\beta}\right)\right).
    \label{eq:poisreg}
\end{equation}
For the linear Poisson regression model in \eqref{eq:poisreg}, it is straightforward to infer the relationship between the covariates and the dependent variable. In practice it is unwise to fit the data with a fully saturated model. A fully saturated model can certainly accounts for all possible interactions between observed covariates in linear form, however it requires to estimate the same number of parameters as the data dimension, which creates extra computational burden and hinders any meaningful interpretation of the results when the dimension becomes large. Including only a limited subset of covariates and their interactions is more feasible, however, the regression can potentially fail to account for residual variation in the observed counts $Y_{i,t,k}$. It may also be at the risk of bias induced by unobserved confounding variables. To address these issues, we propose to combine the current Poisson regression framework with Tensor Train Decomposition technique to form a new Poisson Regression Tensor Train Decomposition (BPRTTD) model so that we are able to extract more information from the data. The model extends the Poisson regression model with an extra rate parameter $\lambda^*_{i,t,k}$
\begin{equation}
    Y_{i,t,k} \sim \text{Pois}\left(u_{i,t,k}\exp\left(\mathbf{x}_{i,t,k}\cdot\boldsymbol{\beta}\right)\lambda^*_{i,t,k}\right).
    \label{eq:BPRTTD}
\end{equation}
We assume that the rate $\lambda^*_{i,t,k}$ can be expressed according to tensor train decomposition such that
\begin{align*}
    \lambda^*_{i,t,k} &= \sum_{h_1=1}^{H_1} \lambda^{(1)}_{i,h_1}\sum_{h_2}^{H_2}\lambda^{(2)}_{t,h_1,h_2}\lambda^{(3)}_{k,h_2} \\
    &=\boldsymbol{\lambda}^{(1)'}_i \Lambda^{(2)}_{t} \boldsymbol{\lambda}^{(3)}_{k},
\end{align*}
where $\boldsymbol{\lambda}^{(1)}_{i}=(\lambda^{(1)}_{i,1},\dots, \lambda^{(1)}_{i,H_1})'\in \mathbb{R}_+^{H_1}, \boldsymbol{\lambda}^{(3)}_{k}=(\lambda^{(3)}_{k,1},\dots, \lambda^{(3)}_{k,H_2})\in\mathbb{R}_+^{H_2}$ and 
\begin{equation*}
    \Lambda^{(2)}_{t}=\begin{pmatrix}
\lambda^{(2)}_{t,1,1} & \lambda^{(2)}_{t,1,2} & \dots & \lambda^{(2)}_{t,1,H_2} \\
\lambda^{(2)}_{t,2,1} & \lambda^{(2)}_{t,2,2} & \dots & \lambda^{(2)}_{t,2,H_2} \\
\vdots & \vdots & \dots & \vdots \\
\lambda^{(2)}_{t,H_1,1} & \lambda^{(2)}_{t,H_1,2} & \dots & \lambda^{(2)}_{t,H_1,H_2}
\end{pmatrix}.
\end{equation*}
Here collection of matrices $\{\boldsymbol{\lambda}^{(1)}_i\}_{i=1,\dots,N}, \{\Lambda_t^{(2)}\}_{t=1,\dots,T}$ and $\{\boldsymbol{\lambda}^{(3)}_k\}_{k=1,\dots,K}$ are tensor train cores.
$H_1$ and $H_2$ are tensor train ranks and they control the model complexity. When $H_1$ and $H_2$ are small relative to $N, T$ and $K$, this is a parsimonious representation of the rate tensor $\{\lambda^*_{i,t,k}\}_{i=1,\dots,N, t=1,\dots,T, k=1,\dots,K}$. Initially, the tensor has $N\cdot T \cdot K$ parameters whereas the number reduces to $N\cdot H_1+T\cdot H_1\cdot  H_2+K\cdot H_2$ after using the tensor decomposition representation.

When the data are Poisson counts and are treated as tensors, \citet{schein2015bayesian} and \citet{schein2016bayesian} applied CP decomposition and Tucker decomposition to enforce dimension reduction and obtain reliable statistical inferences. More recently, tensor train decomposition has gained more popularity. For instance, \citet{mehrizi2021trend} proposed a content request prediction algorithm that employs tensor train decomposition. Motivated by existing literature, our method combines the classical Poisson regression model and the tensor train decomposition to fully utilize information in the data. Furthermore, since we are more oriented in explanatory analysis than predictive performance of the approach, we carefully specify the priors and choose the set of prior hyperparameters to avoid unidentifiable issues inherently to the general latent factor models.

\section{Prior Specification and Posterior Inference}
\label{Posterior}

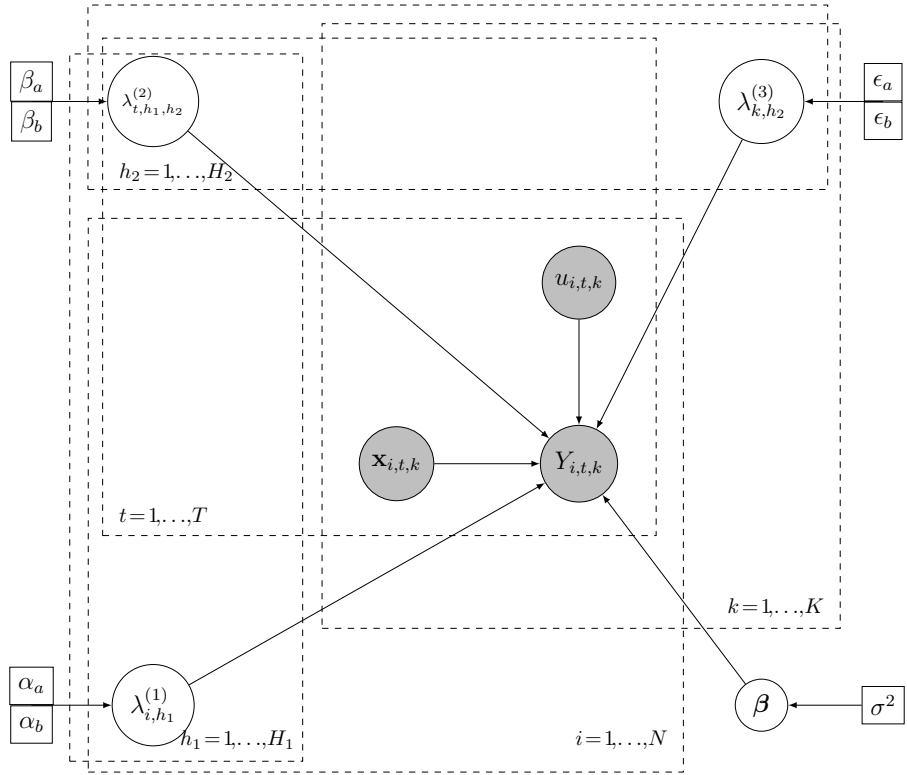
\begin{figure}[t]
\centering
\resizebox{0.75\textwidth}{!}{
\begin{tikzpicture}
\tikzset{box/.style = {rectangle, inner sep=0pt, text width=20mm, text height=-5mm, font=\footnotesize}}
\tikzset{vertex1/.style = {shape=circle,draw,minimum size=1.5em}}
\tikzset{vertex2/.style = {shape=rectangle,draw,minimum size=1.5em}}
\tikzset{vertex3/.style = {shape=rectangle,draw,dashed,minimum size=1.5em}}
\tikzset{vertex4/.style =
{shape=circle,draw,fill=gray!50,minimum size=1.5em}}
\tikzset{edge/.style = {->,> = latex}}
\tikzset{fit margins/.style={/tikz/afit/.cd,#1,/tikz/.cd,inner xsep=\pgfkeysvalueof{/tikz/afit/left}+\pgfkeysvalueof{/tikz/afit/right},inner ysep=\pgfkeysvalueof{/tikz/afit/top}+\pgfkeysvalueof{/tikz/afit/bottom},xshift=-\pgfkeysvalueof{/tikz/afit/left}+\pgfkeysvalueof{/tikz/afit/right},yshift=-\pgfkeysvalueof{/tikz/afit/bottom}+\pgfkeysvalueof{/tikz/afit/top}},afit/.cd,left/.initial=2pt,right/.initial=2pt,bottom/.initial=2pt,top/.initial=2pt}

\node[vertex2] (cons1) at (-9,6.33) {$\beta_a$};
\node[vertex2] (cons2) at (-9,5.67) {$\beta_b$};
\node[vertex2] (cons3) at (5,6.32) {$\epsilon_a$};
\node[vertex2] (cons4) at (5,5.68) {$\epsilon_b$};
\node[vertex2] (cons5) at (-9,-3.68) {$\alpha_a$};
\node[vertex2] (cons6) at (-9,-4.32) {$\alpha_b$};
\node[vertex2] (cons7) at (5,-4) {$\sigma^2$};

\node[vertex1,font=\scriptsize] (par2) at (-7,6) {$\lambda^{(2)}_{t,h_1,h_2}$};
\node[vertex1] (par3) at (3,6) {$\lambda^{(3)}_{k,h_2}$};
\node[vertex1] (par1) at (-7,-4) {$\lambda^{(1)}_{i,h_1}$};
\node[vertex1] (par4) at (3,-4) {$\boldsymbol{\beta}$};

\node[vertex4] (obs1) at (0,0) {$Y_{i,t,k}$};
\node[vertex4] (obs2) at (0,3) {$u_{i,t,k}$};
\node[vertex4] (obs3) at (-3,0) {$\mathbf{x}_{i,t,k}$};

\node[box] (plate2_text) at (-6.5,-1) {$\!t\!=\!1\!,\!\dots\!,\!T$};
\node[vertex3,fit margins={left=9pt,right=1pt,bottom=7pt,top=5pt},fit=(par2) (obs1) (obs2) (obs3)] (plate2) at (-3,3) {};

\node[box] (plate3_text) at (3.5,-2.5) {$\!k\!=\!1\!,\!\dots\!,\!K$};
\node[vertex3,fit margins={left=8pt,right=9pt,bottom=15pt,top=23pt},fit=(par3) (obs1) (obs2) (obs3)] (plate3) at (0,2) {};

\node[box] (plate1_text) at (1,-4.7) {$\!i\!=\!1\!,\!\dots\!,\!N$};
\node[vertex3,fit margins={left=20pt,right=15pt,bottom=19pt,top=7pt},fit=(par3) (obs1) (obs2) (obs3)] (plate1) at (-3,-0.1) {};

\node[box] (plate4_text) at (-6.5,4.7) {$\!h_2\!=\!1\!,\!\dots\!,\!H_2$};
\node[vertex3,fit margins={left=12pt,right=-2pt,bottom=10pt,top=12pt},fit=(par2) (par3)] (plate4) at (-1.5,6) {};

\node[box] (plate5_text) at (-5.5,-4.7) {$\!h_1\!=\!1\!,\!\dots\!,\!H_1$};
\node[vertex3,fit margins={left=23pt,right=10pt,bottom=3pt,top=1pt},fit=(par2) (par1)] (plate5) at (-6,1) {};

\draw[edge] (-9,6) to (par2);
\draw[edge] (5,6) to (par3);
\draw[edge] (-9,-4) to (par1);
\draw[edge] (cons7) to (par4);
\draw[edge] (par1) to (obs1);
\draw[edge] (par2) to (obs1);
\draw[edge] (par3) to (obs1);
\draw[edge] (par4) to (obs1);
\draw[edge] (obs2) to (obs1);
\draw[edge] (obs3) to (obs1);

\end{tikzpicture}
}
\caption{A directed graph summarizing the prior specification of the BPRTTD model. Square boxes are pre-fixed constant hyperparameters; circles are parameters of inferential interest and the colored circles are observed quantities.} \label{fig:hierarchical diagram}
\end{figure}

Due to the complex nature of the model space, we adapt a Bayesian approach to make inferences. Bayesian methods also provide the necessary uncertainty quantification. We impose gamma priors on $\{\boldsymbol{\lambda}^{(1)}_i\}_{i=1,\dots,N}, \{\Lambda_t^{(2)}\}_{t=1,\dots,T}$ and $\{\boldsymbol{\lambda}^{(3)}_k\}_{k=1,\dots,K}$ to exploit the congugate property of the Poisson parameters; that is
\begin{gather*} 
    \lambda^{(1)}_{i,h_1}\sim \text{Ga}(\alpha_a, \alpha_b),\ i=1,\dots,N ,\ h_1=1,\dots,H_1, \\
    \lambda^{(2)}_{t,h_1,h_2} \sim \text{Ga}(\beta_a, \beta_b), \ t=1,\dots,T,\ h_1=1,\dots,H_1,\ h_2=1,\dots,H_2, \\
    \lambda^{(3)}_{k,h_2}\sim \text{Ga}(\epsilon_a, \epsilon_b),\ k=1,\dots,K,\ h_2=1,\dots,H_2.
\end{gather*}
Posterior inference on these parameters can be obtained by using Gibbs sampling algorithm conditionally on most recent values of other parameters. As for the Poisson regression coefficients $\beta$, we follow the literature and assume zero-mean normal priors such that
\begin{equation*}
    \beta_{p}\sim \mathcal{N}(0, \sigma^2),\ p=1,\dots,P.
\end{equation*}
This completes the prior specification for the BPRTTD model. Figure \ref{fig:hierarchical diagram} illustrates the hierarchical graphical representation of the model together with the imposed priors.

Since normal priors on $\boldsymbol{\beta}$ are not conjugate, we sample $\boldsymbol{\beta}$ in an adaptive Metropolis-Hastings step that learns the posterior correlation between multivariate parameters \citep{roberts2009examples}. We outline the MCMC algorithm in below.

\subsection{Metropolis within Gibbs sampler}

We employ a Gibbs sampler for $\lambda_{i,h_1}, \lambda_{t,h_1,h_2}$ and $\lambda_{k,h_2}$ given the Poisson regression coefficients $\boldsymbol{\beta}$. The Gibbs sampling algorithm augments the state space with variable $Y^{h_1,h_2}_{i,t,k}$ such that
\begin{equation}
    Y^{h_1,h_2}_{i,t,k}\sim \text{Pois}\left(u_{i,t,k}\exp\left(\mathbf{x}_{i,t,k}\cdot\boldsymbol{\beta}\right)\lambda^{(1)}_{i,h_1}\lambda^{(2)}_{t,h_1,h_2}\lambda^{(3)}_{k,h_2}\right).
    \label{eq:latent_BPRTTD}
\end{equation}
Utilizing the closeness under addition property of Poisson random variables, \eqref{eq:latent_BPRTTD} implies that
\begin{equation*}
    Y_{i,t,k} = \sum_{h_1=1}^{H_1}\sum_{h_2=1}^{H_2}Y^{h_1,h_2}_{i,t,k}.
\end{equation*}
To draw $Y_{i,t,k}^{h_1,h_2}$ conditional on $Y_{i,t,k}$ and $\lambda^{(1)}_{i,h_1}, \lambda^{(2)}_{t,h_1,h_2}, \lambda^{(3)}_{k,h_2}$, it suffices to note the relationship between the Poisson random variable and the Multinomial random variable, i.e.
\begin{equation*}    \left(Y_{i,t,k}^{1,1},Y_{i,t,k}^{1,2},\dots,Y_{i,t,k}^{H_1,H_2}\right) \sim \text{Multi}\left(Y_{i,t,k}, \left(\pi_{i,t,k}^{1,1},\pi_{i,t,k}^{1,2},\dots,\pi_{i,t,k}^{H_1,H_2}\right)\right)
\end{equation*}
with $\pi_{i,t,k}^{h_1,h_2}=\lambda^{(1)}_{i,h_1}\lambda^{(2)}_{t,h_1,h_2}\lambda^{(3)}_{k,h_2}/\sum_{h_1=1}^{H_1}\sum_{h_2=1}^{H_2}\lambda^{(1)}_{i,h_1}\lambda^{(2)}_{t,h_1,h_2}\lambda^{(3)}_{k,h_2}$. Other useful latent quantities for Gibbs sampler that follows are 
\begin{align*}
    Y^{h_1,\cdot}_{i,\cdot,\cdot} & = \sum_{t=1}^T\sum_{k=1}^K\sum_{h_2=1}^{H_2}Y^{h_1,h_2}_{i,t,k} \\
    &\sim \text{Pois}\left(\lambda^{(1)}_{i,h_1}u_{i,t,k}\exp\left(\mathbf{x}_{i,t,k}\cdot\boldsymbol{\beta}\right)\sum_{t=1}^T\sum_{k=1}^K\sum_{h_2=1}^{H_2}\lambda^{(2)}_{t,h_1,h_2}\lambda^{(3)}_{k,h_2}\right), \\
    Y^{h_1,h_2}_{\cdot,t,\cdot} & = \sum_{i=1}^N\sum_{k=1}^K Y^{h_1,h_2}_{i,t,k} \\
    & \sim \text{Pois}\left(\lambda^{(2)}_{t,h_1,h_2}u_{i,t,k}\exp\left(\mathbf{x}_{i,t,k}\cdot\boldsymbol{\beta}\right)\sum_{i=1}^N\sum_{k=1}^K \lambda^{(1)}_{i,h_1}\lambda^{(3)}_{k,h_2}\right), \\
    Y^{\cdot,h_2}_{\cdot,\cdot,k} & = \sum_{i=1}^N\sum_{t=1}^T\sum_{h_1=1}^{H_1} Y^{h_1,h_2}_{i,t,k} \\
    & \sim \text{Pois}\left(\lambda^{(3)}_{k,h_2}u_{i,t,k}\exp\left(\mathbf{x}_{i,t,k}\cdot\boldsymbol{\beta}\right)\sum_{i=1}^N\sum_{t=1}^T\sum_{h_1=1}^{H_1}\lambda^{(1)}_{i,h_1}\lambda^{(2)}_{t,h_1,h_2}\right).
\end{align*}
With these three auxiliary variables, it is easy to derive the full conditional distributions. To update $\lambda_{i,h_1}$, we draw samples from 
\begin{equation*}
    \lambda_{i,h_1 \mid \cdot} \sim \text{Ga}\left(\alpha_a+Y^{h_1,\cdot}_{i,\cdot,\cdot}, \alpha_b+u_{i,t,k}\exp\left(\mathbf{x}_{i,t,k}\cdot\boldsymbol{\beta}\right)\sum_{t=1}^T\sum_{k=1}^K\sum_{h_2=1}^{H_2}\lambda^{(2)}_{t,h_1,h_2}\lambda^{(3)}_{k,h_2}\right).
\end{equation*}
Similarly for $\lambda_{t,h_1,h_2}$ and $\lambda_{k,h_2}$, the full conditional distributions are
\begin{gather*}
    \lambda_{t,h_1,h_2 \mid \cdot} \sim \text{Ga}\left(\beta_a+ Y^{h_1,h_2}_{\cdot,t,\cdot}, \beta_b+u_{i,t,k}\exp\left(\mathbf{x}_{i,t,k}\cdot\boldsymbol{\beta}\right)\sum_{i=1}^N\sum_{k=1}^K \lambda^{(1)}_{i,h_1}\lambda^{(3)}_{k,h_2}\right) \\
        \lambda_{k,h_2 \mid \cdot} \sim \text{Ga}\left(\epsilon_a+ Y^{\cdot,h_2}_{\cdot,\cdot,k}, \epsilon_b+u_{i,t,k}\exp\left(\mathbf{x}_{i,t,k}\cdot\boldsymbol{\beta}\right)\sum_{i=1}^N\sum_{t=1}^T\sum_{h_1=1}^{H_1}\lambda^{(1)}_{i,h_1}\lambda^{(2)}_{t,h_1,h_2}\right).
\end{gather*}

After updating $\lambda_{i,h_1}, \lambda_{t,h_1,h_2}$ and $\lambda_{k,h_2}$ in each iteration, $\boldsymbol{\beta}$ is sampled in a Metropolis-Hastings step with $n$-step proposal distribution
\begin{equation*}
    Q_n\left(\boldsymbol{\beta},\cdot\right) = (1-p)\mathcal{N}\left(\boldsymbol{\beta}, (2.38)^2\Sigma_n/d \right) + p\mathcal{N}\left(\boldsymbol{\beta}, (0.1)^2\Sigma/d\right),
\end{equation*}
where $p$ is a small constant between 0 and 1, $\Sigma_n$ is empirical estimate of the covariance matrix of the target posterior distribution based on the run so far and $d$ is the dimension of $\boldsymbol{\beta}$. $\Sigma$ is a fixed covariance matrix and we take it to be the GLM estimate of the Poisson regression covariance matrix for efficiency.

\section{Simulation Studies}
\label{Simulation Studies}

We conduct two simulation studies to validate the BPRTTD model and the posterior sampling algorithm. In the first simulation study, we artificially simulate true parameters and use these parameters to generate the Poisson observations. Results are reported in the \nameref{Appendices}. However, the dimension of the simulated data is much smaller than what we encounter in the real data application (see Section \ref{Real Data Application} for more detailed data description). The reason for this choice is that we are able to repeat the simulations multiple times. Another limitation is that the true parameters $\boldsymbol{\beta}$ is sampled from a arbitrary normal distribution, and $ \boldsymbol{\lambda}_i^{(1)}, i=1,\dots,N, \Lambda_t^{(2)}, t=1,\dots,T, \boldsymbol{\lambda}_k^{(3)}, k=1,\dots,K$ are simulated from a gamma distribution with certain artificial shape and rate, which may not really reflect the typical real data scenario. To address these drawbacks, we design the second simulation study where true parameters are estimated from the real data under the BPRTTD model specification (see Section \ref{Real Data Application} for steps regarding posterior inferences). After obtaining the estimates, which we treat as true parameters, we simulate offset from a gamma distribution with shape equal to $10^6$ and rate equal to 1. The Poisson observations are then sampled according to \eqref{eq:BPRTTD}. We apply our approach to the simulated data and verify whether we are able to recover the true parameters in this high-dimensional and more realistic scenario. We report the summary statistics of absolute percentage error (APE) between the true parameters and their posterior mean estimates in Table \ref{tb:sim2}. At least 75\% of parameters $\boldsymbol{\beta}, \lambda_{i,h_1}^{(1)}, \lambda_{t,h_1,h_2}^{(2)}$ and $\lambda_{k,h_2}^{(3)}$ in the BPRTTD model are recovered within 40\% deviation from the truth using our approach.

\begin{table}[ht]
\centering
\resizebox{\textwidth}{!}{
\begin{tabular}{ccccccc}
\hline
 & Min. & 1st Qu. & Median & Mean & 3rd Qu. & Max. \\ \hline
$|\hat{\boldsymbol{\beta}}-\boldsymbol{\beta}|/|\boldsymbol{\beta}|$ & 0.0001 & 0.0281 & 0.0621 & 0.4092 & 0.2152 & 17.5434 \\ \hline
$|\hat{\lambda}^{(1)}_{i,h_1}-\lambda^{(1)}_{i,h_1}|/|\lambda^{(1)}_{i,h_1}|$ & 0.0003 & 0.0765 & 0.1765 & 0.4100 & 0.3681 & 36.9590 \\ \hline
$|\hat{\lambda}^{(2)}_{t,h_1,h_2}-\lambda^{(2)}_{t,h_1,h_2}|/|\lambda^{(2)}_{t,h_1,h_2}|$ & 0.0000 & 0.0864 & 0.1896 & 0.3057 & 0.3818 & 4.1056  \\ \hline
$|\hat{\lambda}^{(3)}_{k,h_2}-\lambda^{(3)}_{k,h_2}|/|\lambda^{(3)}_{k,h_2}|$ & 0.0005 & 0.0342 & 0.0764 & 0.1891 & 0.1978 & 3.1954  \\ \hline
\end{tabular}
}
\caption{Summary statistics of APE between true parameters and the estimated posterior means in the second simulation study.}
\label{tb:sim2}
\end{table}

\section{Drivers of Causes of Death in Italy from 2015 to 2020}
\label{Real Data Application}

With the aim of understanding the shifting mortality patterns of COVID-19 as well as other causes of death prior to and during the pandemic outbreak, we apply our method to Italian official mortality data that records provisional monthly death counts based on the analysis of the declarations of the $K=18$ causes of death compiled by doctors for all deaths in Italy from January 2015 until December 2020, i.e., $T=72$ monthly death counts. Table \ref{tb:death_causes} shows the 18 causes of death under investigation. Furthermore, the death counts are aggregated in $N=420$ levels formed by 10 age groups, 2 genders and 21 Italian regions. In summary, we observe $Y_{i,t,k}$ for $i=1,\dots,N, t=1,\dots,T, k=1,\dots,K$, in total 544,320 observations arrange in a $N\times T\times K$ multiway-array. A more comprehensive description of the mortality data can be found on https://www.istat.it/it/archivio/240401.

\begin{table}[ht]
\centering
\begin{tabular}{ll}
\hline
1.&COVID-19 \\ \hline
2.&Some infectious and parasitic diseases \\ \hline
3.&Tumors \\ \hline
4.&Diseases of the blood and hematopoietic organs and \\
&some disorders of the immune system \\ \hline
5.&Endocrine, nutritional and metabolic diseases \\ \hline
6.&Psychic and behavioral disorders \\ \hline
7.&Diseases of the nervous system and sense organs \\ \hline
8.&Diseases of the circulatory system \\ \hline
9.&Diseases of the respiratory system \\ \hline
10.&Diseases of the digestive system \\ \hline
11.&Diseases of the skin and subcutaneous tissue \\ \hline
12.&Diseases of the musculoskeletal system and connective tissue \\ \hline
13.&Diseases of the genitourinary system \\ \hline
14.&Complications of pregnancy, childbirth and the puerperium \\ \hline
15.&Some morbid conditions that originate in the perinatal period \\ \hline
16.&Congenital malformations and chromosomal anomalies \\ \hline
17.&Symptoms, signs, abnormal results and ill-defined causes \\ \hline
18.&External causes of trauma and poisoning \\ \hline
\end{tabular}
\caption{Causes of death according to the ICD-10.}
\label{tb:death_causes}
\end{table}

Along with death counts $Y_{i,t,k}$, we also obtain covariates $\mathbf{x}_{i,t,k}$. One important variable is the Italian Stringency Index (ISI) presented by \citet{conteduca2022new} in the same spirit as the Oxford Stringency Index (OSI) introduced by \citet{hale2021global}. The data set measures non-pharmaceutical interventions adopted by Italian authorities to tackle the COVID-19 pandemic at both the national and regional levels. Regional level stringency indices are desirable since mortality counts are collected according to regions. We look into interactions between the ISI and various causes of death, as suggested in literature that the pandemic can potentially result in differential consequences in other mortality causes. The other two groups of covariates that we include are interactions between age groups and causes of death as well as interactions between age groups and gender. It is well documented that age and gender are important risk factors for many causes of death. Female and male also demonstrate varying mortality patterns in different ages. These interaction terms in total result in 208 dimensional covariates $\mathbf{x}_{i,t,k}$ in the model. Lastly, the offsets $u_{i,t,j}$ we include are days in each month, the reported monthly aggregated cases in each region for COVID-19 death category and population in each region for all other causes of death. Specifically for external causes of trauma and poisoning, we consider another offset that reflects the mobility level. The index we adapt is the Google COVID-19 Community Mobility Reports (Google LLC "Google COVID-19 Community Mobility Reports". https://www.google.com/covid19/mobility/). By adding the mobility offset into the Poisson rate, we model the change in mortality rate of external causes of death per fixed mobility unit. The remaining Poisson rate $\lambda^*_{i,t,k}$ unaccounted for by the regression component is assumed to be has latent structure with $H_1=6$ and $H_2=6$. The choice of these two values is tested over varying combinations of $H_1$ and $H_2$ over grids defined by $H_1=5,6,7,8$ and $H_2=5,6,7,8$ and we use $H_1=6$ and $H_2=6$ to achieve the balance between reasonable model fitting and model complexity. The Gamma priors on $\lambda^{(1)}_{i,h_1}, \lambda^{(2)}_{t,h_1,h_2}, \lambda^{(3)}_{k,h_2}$ has parameters such that $\alpha_2 = 20, \alpha_1 = \sqrt{1/(H1*H2)}*\alpha_2, 
\beta_2 = 20, \beta_1 = \sqrt{1/(H1*H2)}*\beta_2, \epsilon_1 = 200, \epsilon_2 = 200$. For the Poisson regression coefficients $\boldsymbol{\beta}$, we impose centered normal priors with variance 2. The MCMC iterations are 40,000.



\subsection{Improvement of the BPRTTD model over the Poisson regression}

\begin{figure}[ht]\captionsetup[subfigure]{font=footnotesize}
     \centering
     \resizebox{\textwidth}{!}{
         \begin{subfigure}[t]{0.32\textwidth}
         \centering
         \includegraphics[width=\textwidth]{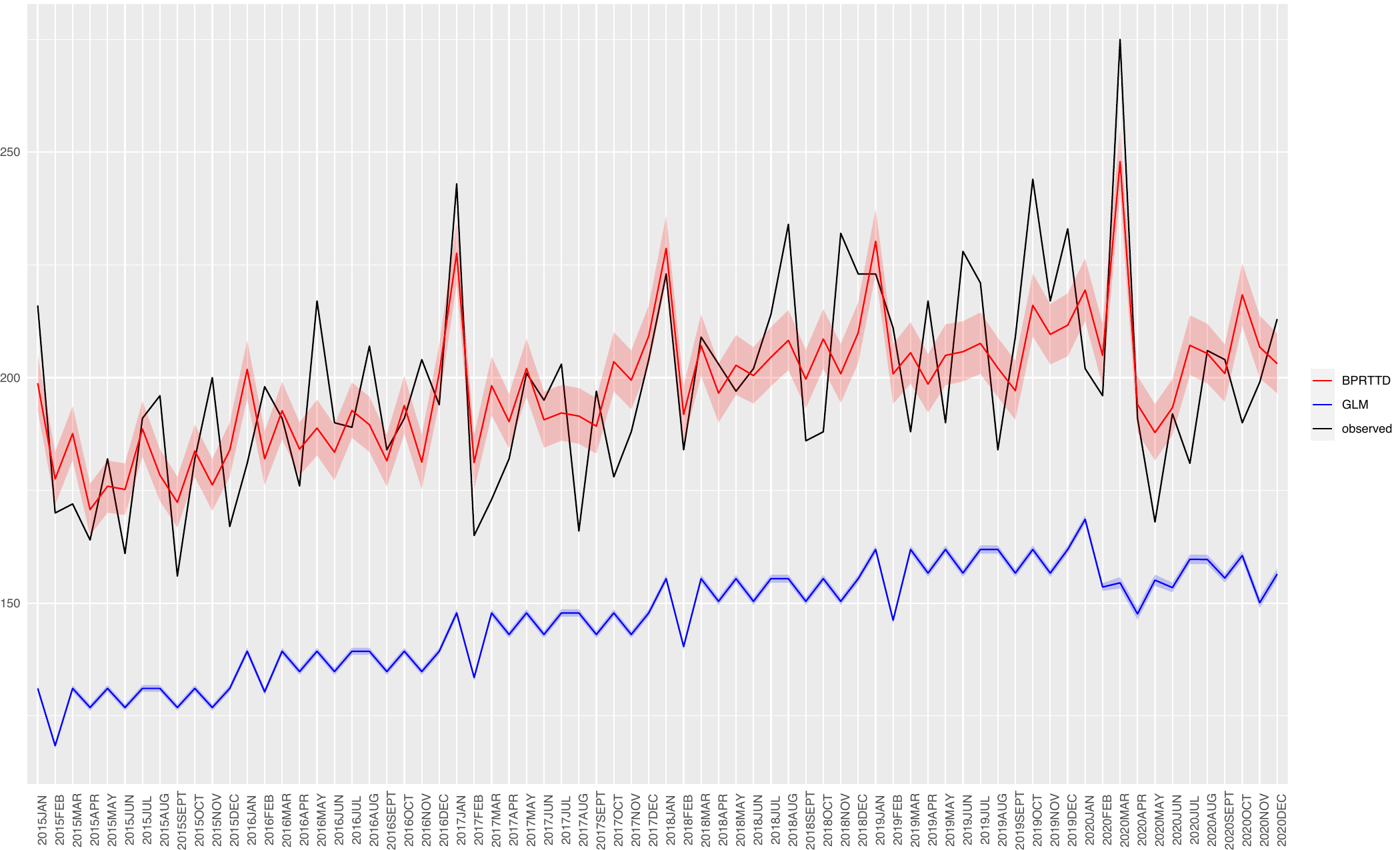}
         \caption{\footnotesize Lombardia, male, 80-84, Tumors}
         \label{subfig:traj1}
     \end{subfigure}
         \begin{subfigure}[t]{0.32\textwidth}
         \centering
         \includegraphics[width=\textwidth]{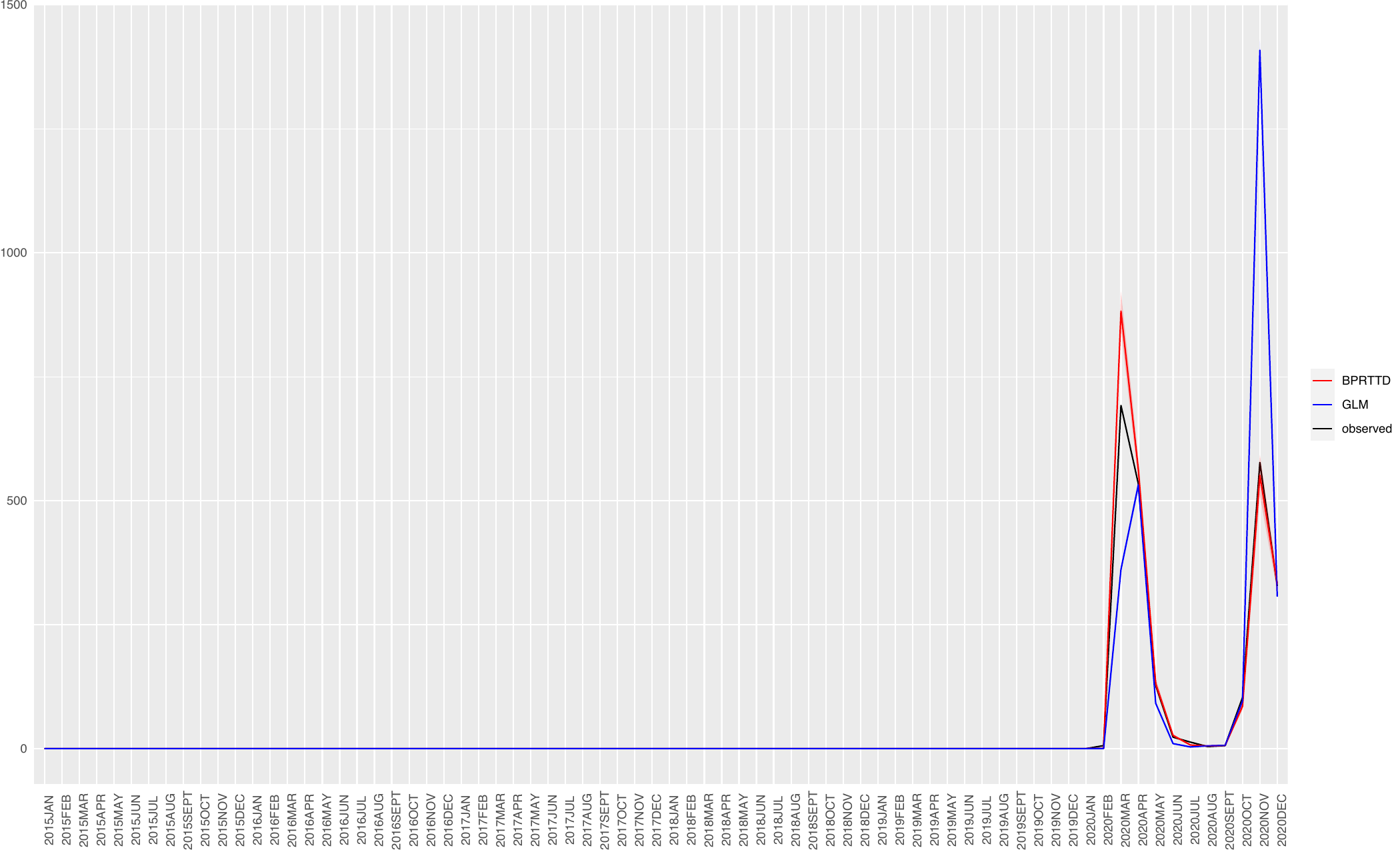}
         \caption{\footnotesize Lombardia, male, 80-84, COVID-19}
         \label{subfig:traj2}
     \end{subfigure}
         \begin{subfigure}[t]{0.32\textwidth}
         \centering
         \includegraphics[width=\textwidth]{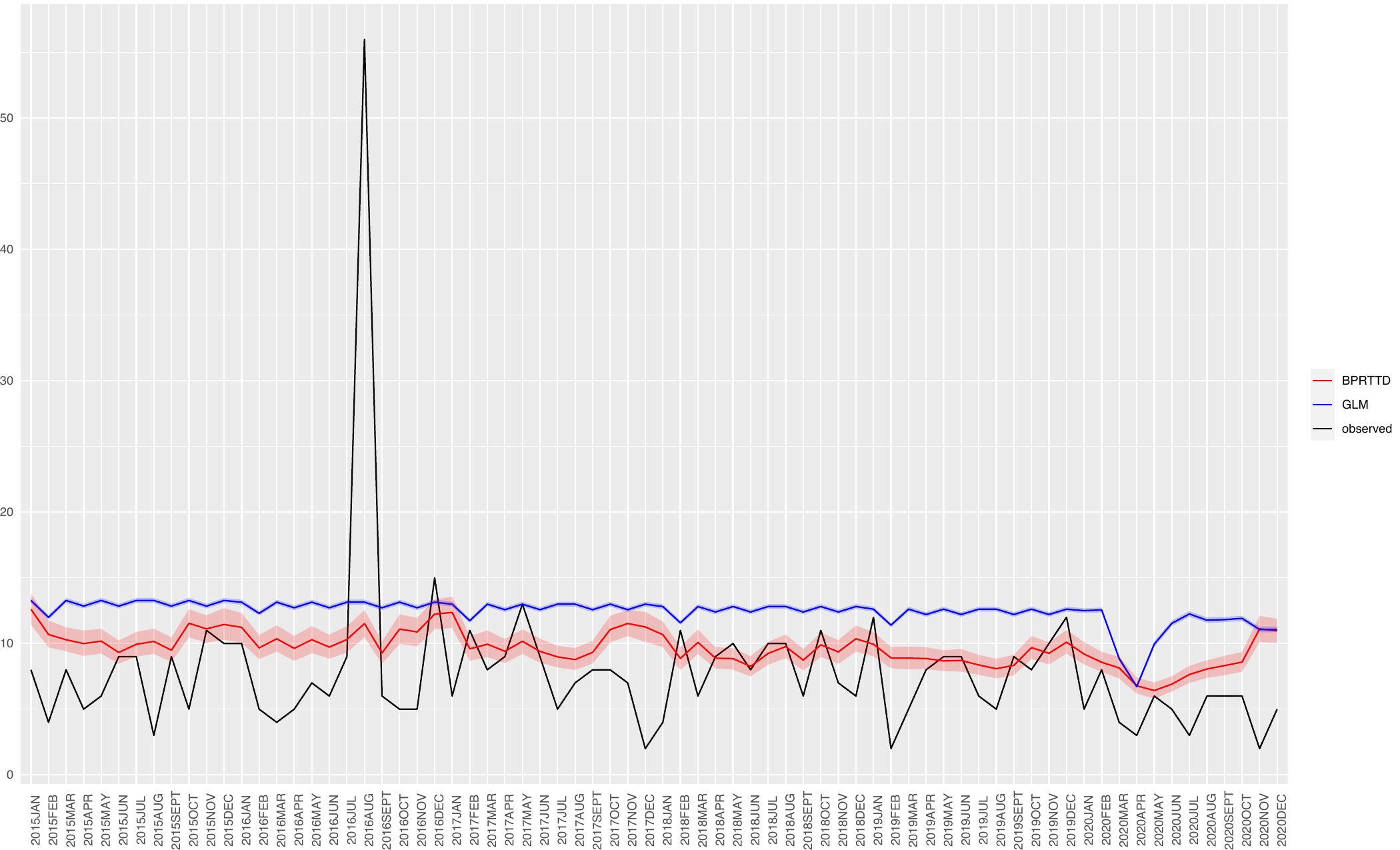}
         \caption{\footnotesize Lazio, female, 0-49, External causes of trauma and poisoning}
         \label{subfig:traj3_2}
     \end{subfigure}
     }
     \caption{Death counts of selected demographic groups and causes of death from January 2015 to December 2020. Black line is the observed trajectory $Y_{i,t,k}, t=1,\dots,T$ for fixed $i$ and $k$; red and blue lines are BPRTTD fitted values and GLM fitted values respectively. Shaded areas correspond to 95\% credible intervals for BPRTTD predicts and 95\% confidence interval for GLM predicts.}
     \label{fig:traj}
\end{figure}

First, we highlight what the additional tensor decomposition component contributes to fitting the Poisson rate by showing in Figure \ref{fig:traj} how our method complements the GLM estimates in recovering the observed variations in death counts $Y_{i,t,k}$. In these selected trajectories, we can see that the tensor decomposition component adjusts the naive GLM estimates to better follow the observed trajectories. For instance, GLM predicted values of death counts of male who reside in Lombardia and died of Tumors between age 80 to 84 are consistently under the observed ones; this is not surprising since GLM tends to estimate and fit with the average of all observations whereas Lombardia, as the most populated region in Italy, has in general larger values of death counts. Our method successfully makes up the gap between data and GLM estimates by amplifying the Poisson rates, as shown in Figure \ref{subfig:traj1}. In the case such as in Figure \ref{subfig:traj2} where the GLM estimates over predict, $\lambda^*_{i,t,k}$ plays the role of downsizing the Poisson rate. Through the tensor decomposition assumption, such adjustments are done in a parsimonious manner. Recall that the saturated model requires in total 544,320 parameters whereas now except for the 208 coefficients, we add only $N\times H_1+T\times H_1\times H_2+K\times H_2=5,220$ more parameters to achieve great improvement in terms of model fitting. This advantage can also be seen when we calculate and compare the log-likelihood of simple Poisson regression versus our BPRTTD model, which are -862910.4 and -731919.9 respectively. Even though our approach provides further approximation to observations, it is still robust to outliers or abnormal records as the model specification exploits and leverages information from other data by introducing common shared latent classes. We demonstrate in Figure \ref{subfig:traj3_2} such a scenario where female mortality counts in age group 0-49 in Lazio in August 2016 show a sudden spike deviating from the normal pattern. The red BPRTTD line is not sensitive to such an outlier.

\subsection{Interpretation of the Poisson regression component}

\begin{figure}[ht]\captionsetup[subfigure]{font=scriptsize}
     \centering
     
         \begin{subfigure}[t]{0.32\textwidth}
         \centering
         \includegraphics[width=\textwidth]{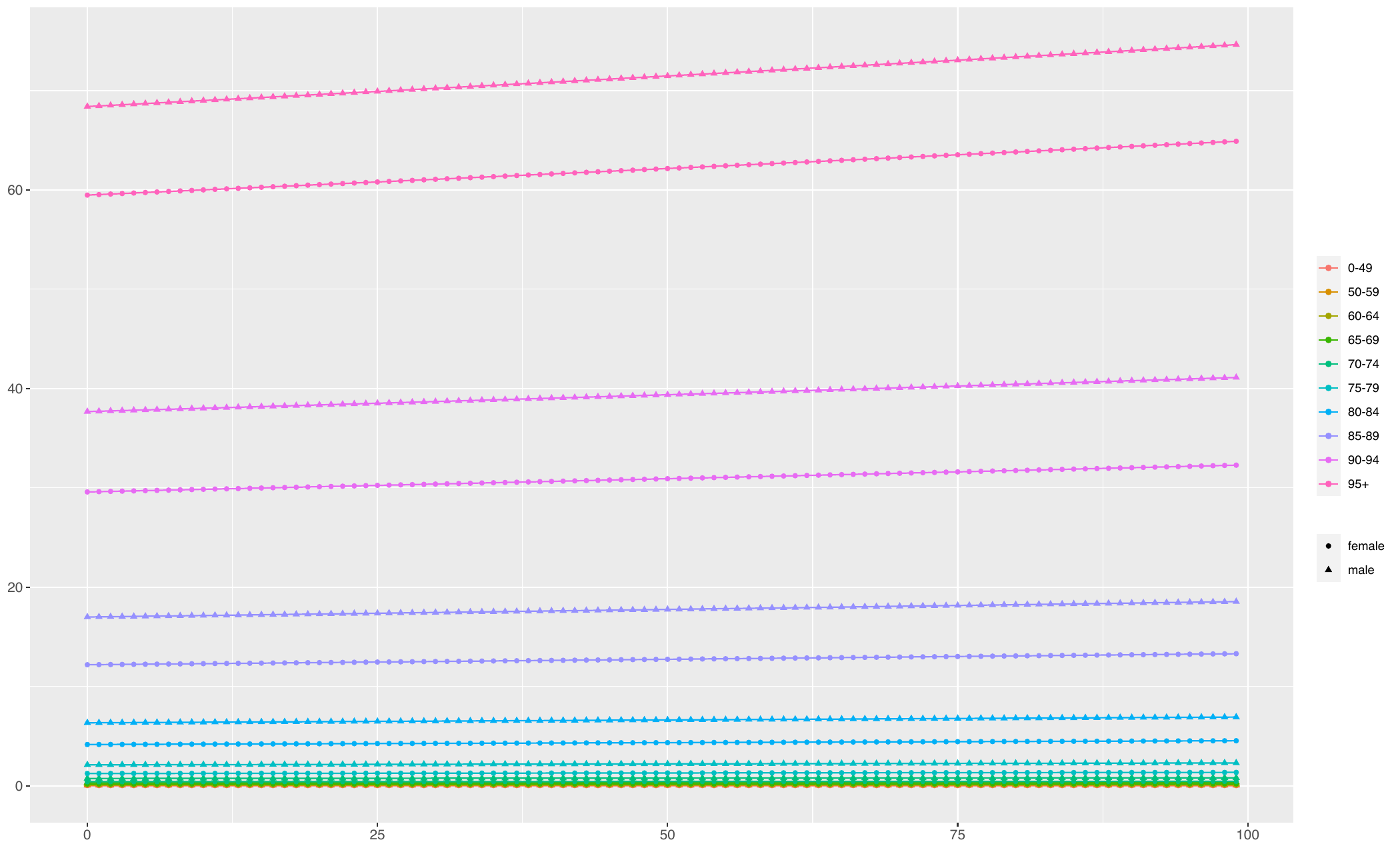}
         \caption{\scriptsize psychic and behavioral disorders}
         \label{subfig:BPRTTD_pred6}
     \end{subfigure}
         \begin{subfigure}[t]{0.32\textwidth}
         \centering
         \includegraphics[width=\textwidth]{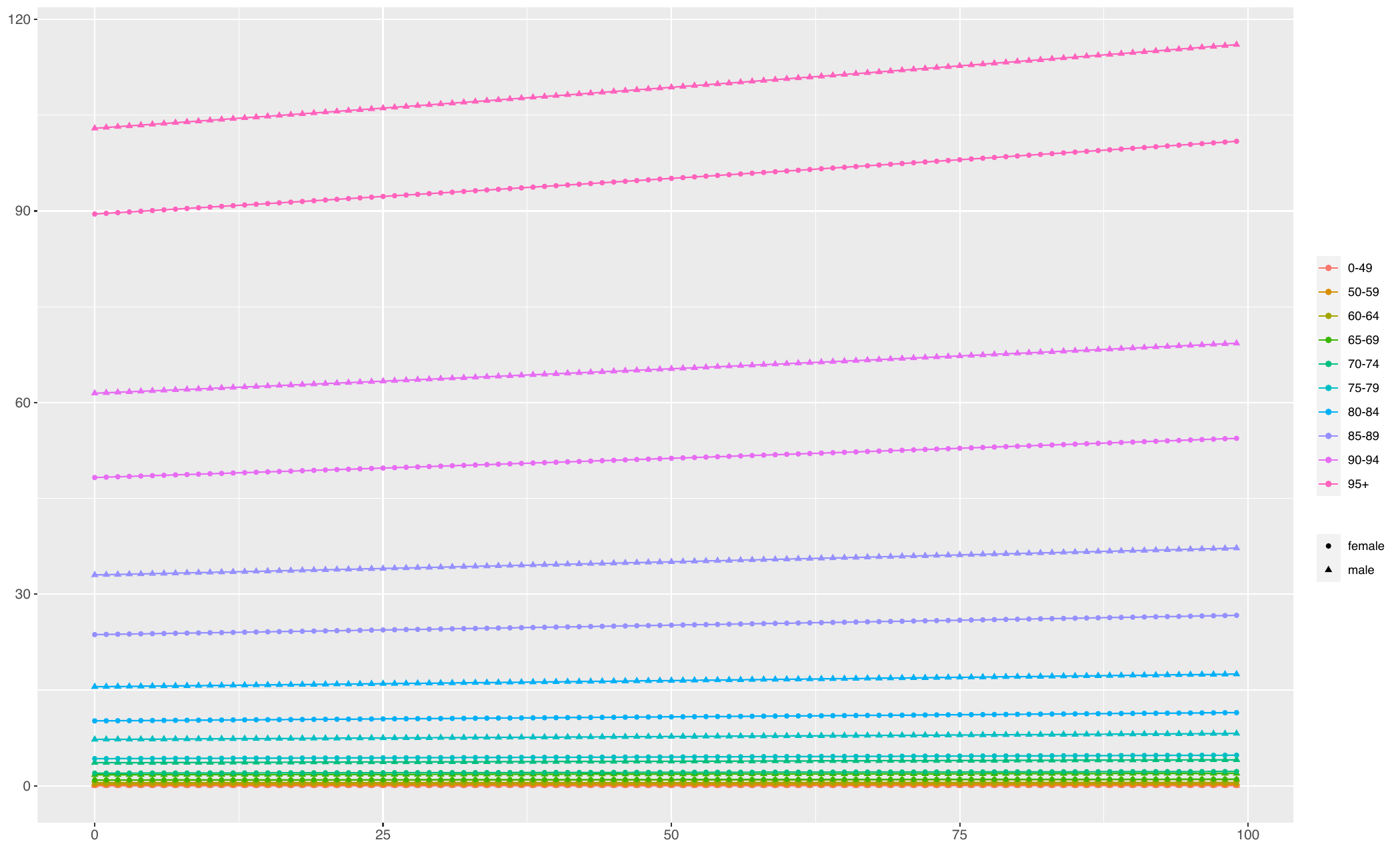}
         \caption{\scriptsize diseases of the respiratory system}
         \label{subfig:BPRTTD_pred9}
     \end{subfigure}
         \begin{subfigure}[t]{0.32\textwidth}
         \centering
         \includegraphics[width=\textwidth]{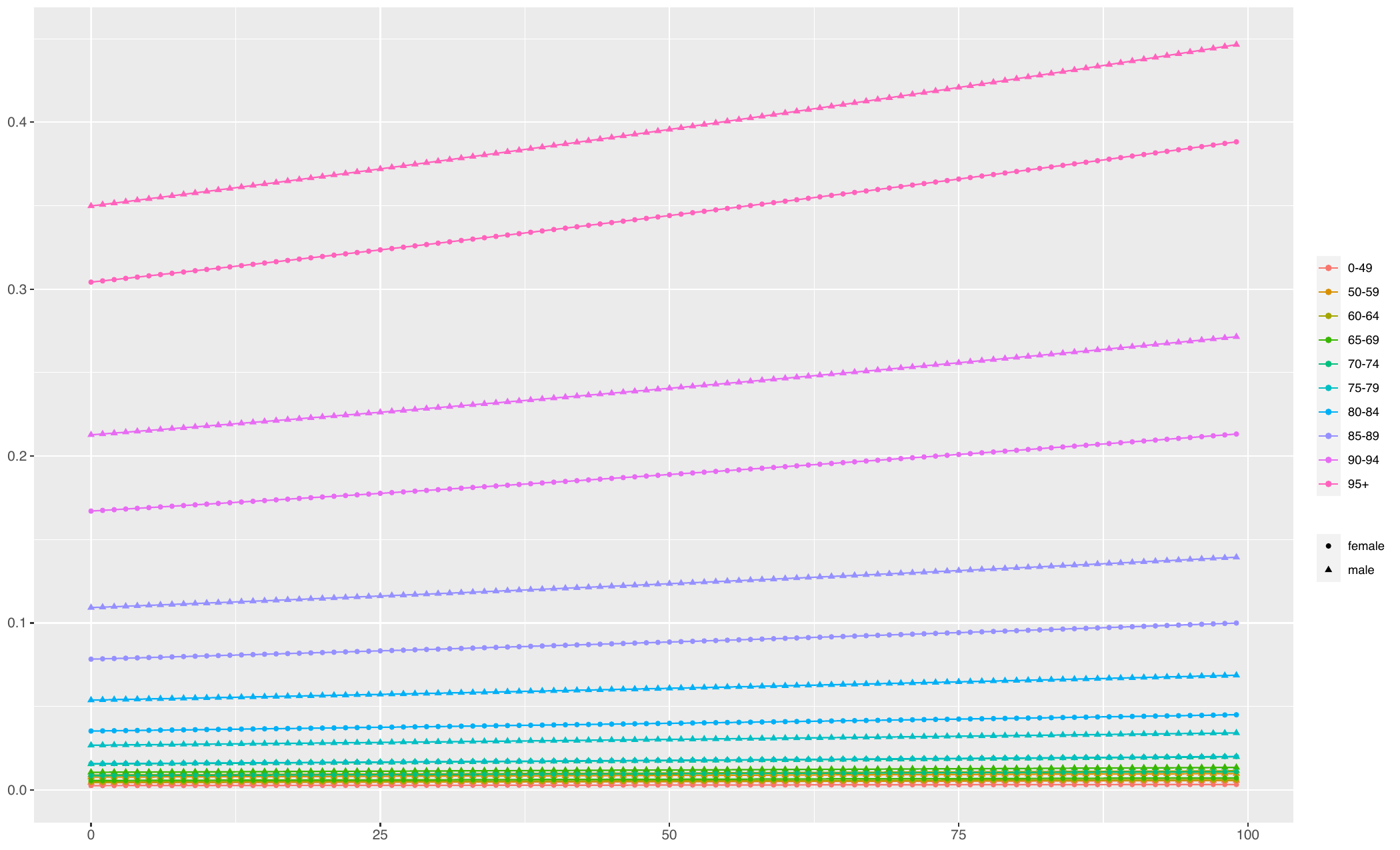}
         \caption{\scriptsize external causes of trauma and poisoning}
         \label{subfig:BPRTTD_pred18}
     \end{subfigure}
              \begin{subfigure}[t]{0.32\textwidth}
         \centering
         \includegraphics[width=\textwidth]{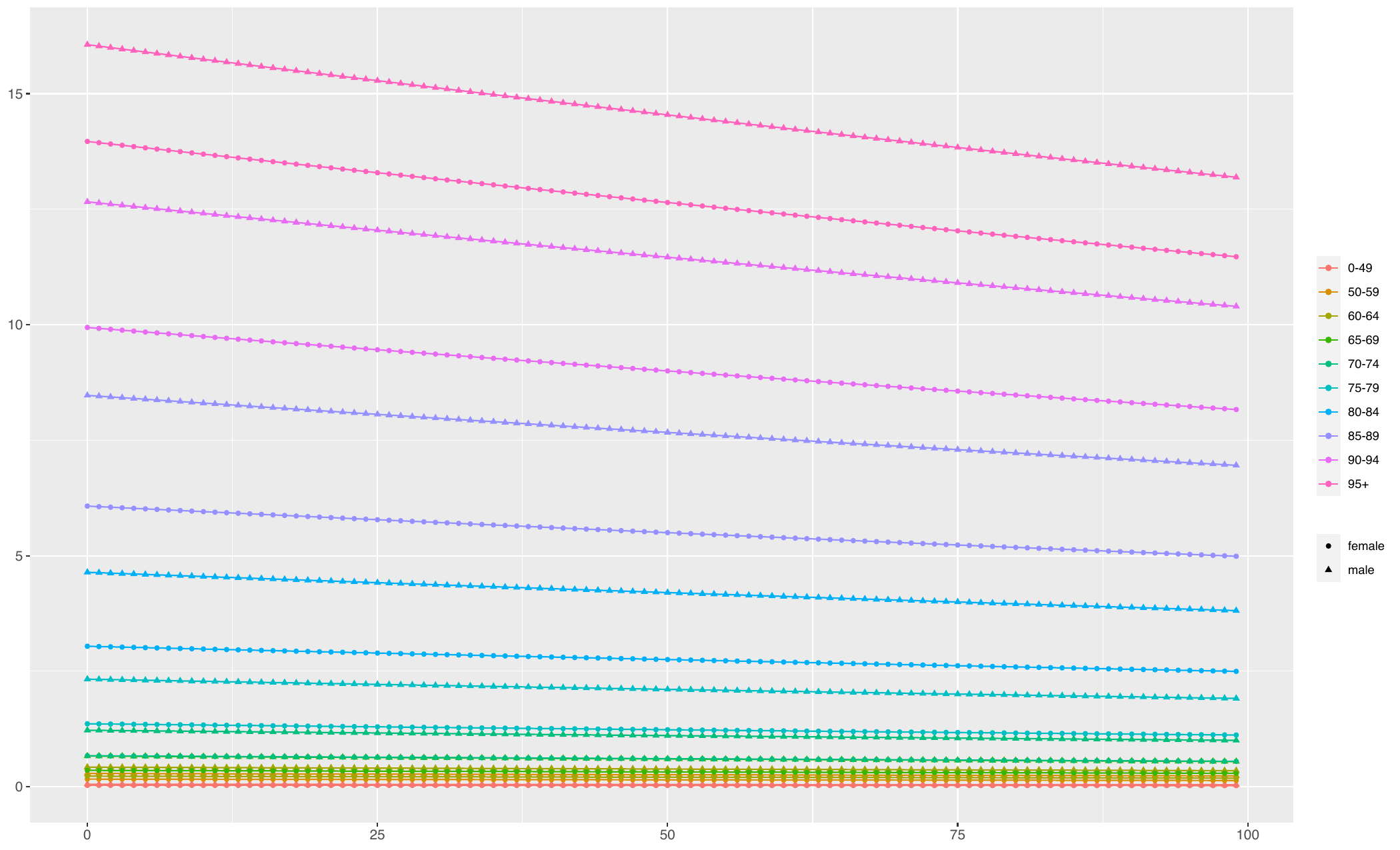}
         \caption{\scriptsize some infectious and parasitic diseases}
         \label{subfig:BPRTTD_pred2}
     \end{subfigure}
         \begin{subfigure}[t]{0.32\textwidth}
         \centering
         \includegraphics[width=\textwidth]{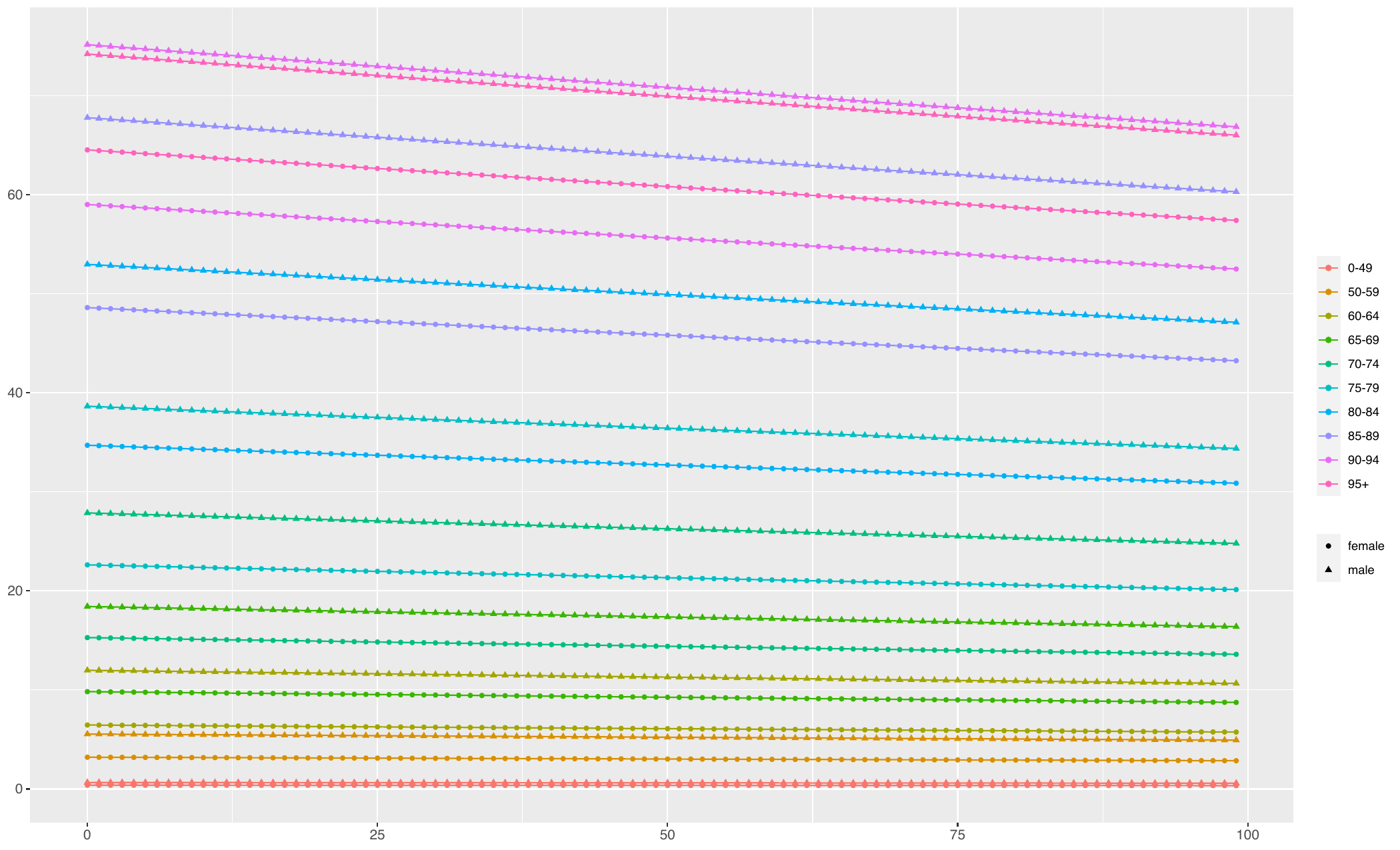}
         \caption{\scriptsize tumors}
         \label{subfig:BPRTTD_pred3}
     \end{subfigure}
         \begin{subfigure}[t]{0.32\textwidth}
         \centering
         \includegraphics[width=\textwidth]{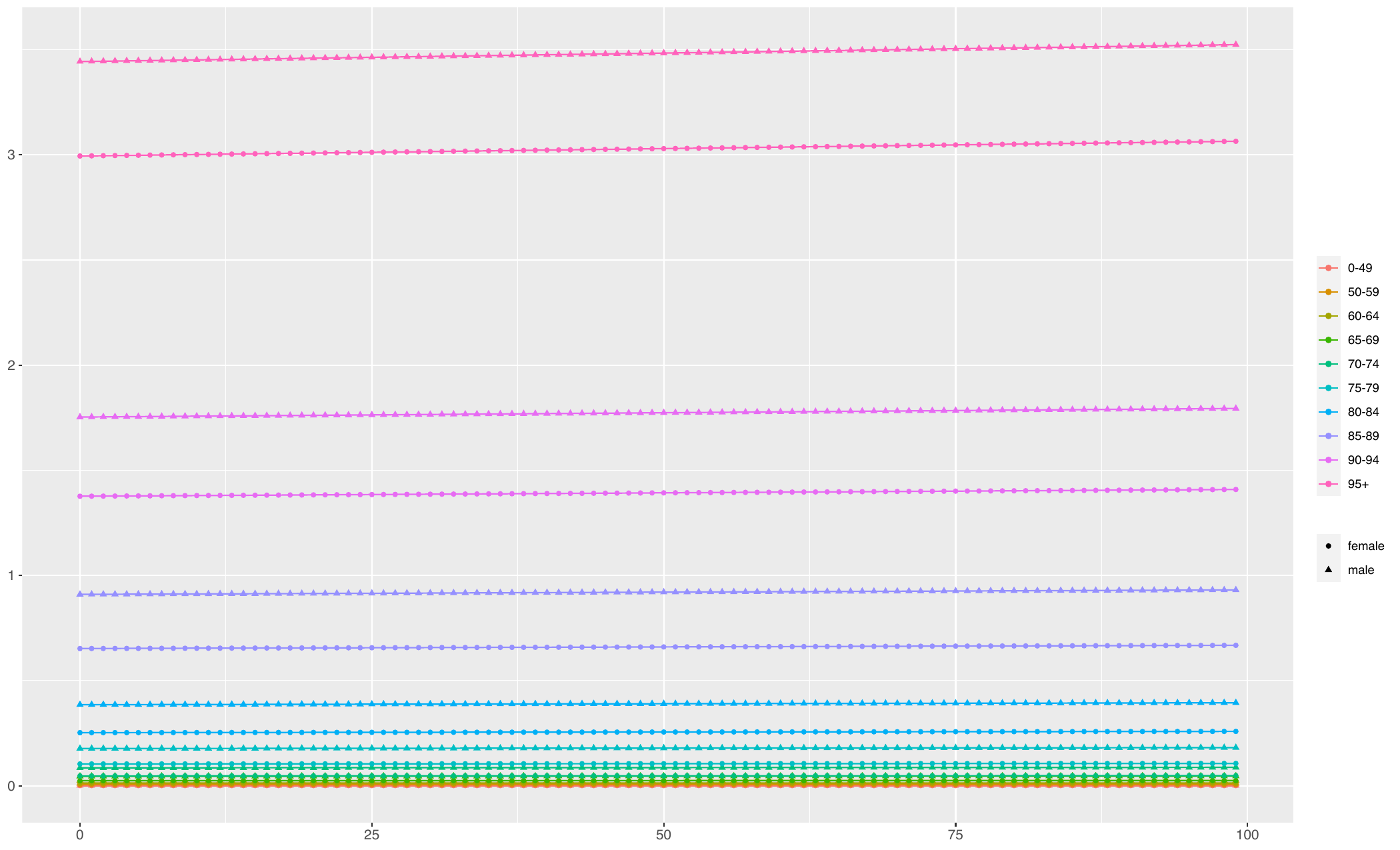}
         \caption{\scriptsize diseases of the skin and subcutaneous tissue}
         \label{subfig:BPRTTD_pred11}
     \end{subfigure}
     \caption{Selected predicted values of the BPRTTD model based on regression coefficients. Horizontal axes are the ISI from 0 to 100 and vertical axes are hazard rates. }
     \label{fig:BPRTTD_pred}
\end{figure}

We now make explanatory analysis on the Italian mortality data. We are primarily interested in discovering how other causes of death are affected by the government intervention policies. Three types of responses are inferred, positive, negative and no effects based on the criteria whether the 95\% credible intervals of each coefficient are above 0, below 0 or contain 0. Mortality counts are positively associated with the ISI in the following death categories: 4. Diseases of the blood and hematopoietic organs and
some disorders of the immune system, 5. Endocrine, nutritional and metabolic diseases, 6. Psychic and behavioral disorders, 7. Diseases of the nervous system and sense organs, 9. Diseases of the respiratory system, 12. Diseases of the musculoskeletal system and connective tissue, 13. Diseases of the genitourinary system, 17. Symptoms, signs, abnormal results and ill-defined causes, 18. External causes of trauma and poisoning. The positive relationship between psychic and behavioral disorders, shown in Figure \ref{subfig:BPRTTD_pred6}, is well documented in literature, affecting psychiatric patients as well as health population \citep{hao2020psychiatric, every2020psychological, pieh2021mental, rossi2020covid}. However, most studies report increasing levels of anxiety, acute stress disorders and so on, we offer new evidence that it actually translates to elevated mortality rates of psychic and behavioral disorders in the end. During the pandemic, individuals with psychiatric and behavioral disorders may face additional challenges due to disruptions in routine care, limited access to mental health services, increased stressors and social isolation. These factors can potentially contribute to adverse outcomes and exacerbate existing conditions. Another positive relationship we would like to comment on is between mortality due to respiratory system diseases and the ISI in Figure \ref{subfig:BPRTTD_pred9}. Although there have been wide range of studies suggesting that people with certain lung diseases appear to have an increased risk at the height of the epidemic and these risk factors are important clinical predictors of severe COVID-19 to enable risk stratification and optimize resource allocation \citep{lippi2020chronic, aveyard2021association}, we discover that reversely the mortality rate of respiratory disease rises during COVID-19 lockdown despite the common observation that respiratory disease incidences declined due to public precautionary measures \citep{hsieh2020outcome, huh2021decrease, britton2020covid}. Several factors can jointly explain the positive relationship. For instance, lockdown measures can disrupt the routine care and monitoring of respiratory conditions, as a result, lack of timely interventions and preventive measures can contribute to a higher risk of mortality. Misclassification can also explain the increasing mortality. In the early pandemic, diagnosing the cause of death accurately can be complex especially when healthcare systems are under strain; limited testing capacity or availability of COVID-19 tests also potentialize deaths being attributed to respiratory diseases without confirming the presence of COVID-19. As for the mortality rate of external causes of trauma and poisoning in Figure \ref{subfig:BPRTTD_pred18}, it may be contradicting to see that this also trended up as more intense lockdown measures were enforced. However, since we include mobility index in the offset that disentangles the negative effect of lockdown on population mobility from the total effect, we state that government intervention policies actually drive up mortality per mobility unit due to reasons such as delayed or reduced access to healthcare.

Negative correlations appear in 2. Some infectious and parasitic diseases, see Figure \ref{subfig:BPRTTD_pred2}, 3. Tumors, see Figure \ref{subfig:BPRTTD_pred3}, 8. Diseases of the circulatory system, 10. Diseases of the digestive system, 14. Complications of pregnancy, childbirth and the puerperium, 15. Some morbid conditions that originate in the perinatal period, 16. Congenital malformations and chromosomal anomalies. It has been observed that infectious and parasitic diseases caused less mortality when government interventions were more strict \citep{dadras2021effects}. Measures such as lockdowns, travel restrictions, and social distancing, can help limit the spread of infectious diseases. By reducing contact between individuals, these measures can interrupt the transmission of infectious agents, thereby decreasing the overall incidence of infections and subsequent mortality. As for the decrease in tumor mortality rate, one possible explanation is the harvesting effect, also known as mortality displacement \citep{schwartz2000harvesting, kepp2022estimates}. The harvesting effect refers to the phenomenon that individuals who are already vulnerable, in this case, tumor patients, experience accelerated deaths during the COVID-19 lockdown intervention, leading to a temporary decline in tumor mortality rates. However, this decline is expected to be followed by a period of increased mortality as those who would have died during the intervention succumb in the subsequent period. Figure \ref{subfig:BPRTTD_pred11} shows the only category 11. Diseases of the skin and subcutaneous tissue that exhibits no statistically significant relationship with ISI. In Figure \ref{fig:BPRTTD_pred}, we can also conclude the effect of gender and age on the hazard rates. In general, older population is associated with higher mortality in almost all types of death causes and men are more likely to die than women in the same age group. The exception is with tumors where men from certain younger age groups present higher mortality rates compared to women from older age group. It is also counter-intuitive to observe that the mortality rate due to external causes of trauma and poisoning is positively related to age. Even though it is confirmed in the data that the absolute death counts do go down with age, after taking into account the population size of each age group, the mortality rates per unit of population indeed increase with age, indicating that external causes of trauma and poisoning becomes more threatening when people get older. For detailed coefficient estimates, please refer to the \nameref{Appendices}.
\subsection{Interpretation of the latent parameters}

\begin{figure}
     \centering
    \begin{subfigure}[b]{0.47\textwidth}
         \centering
         \includegraphics[width=\textwidth]{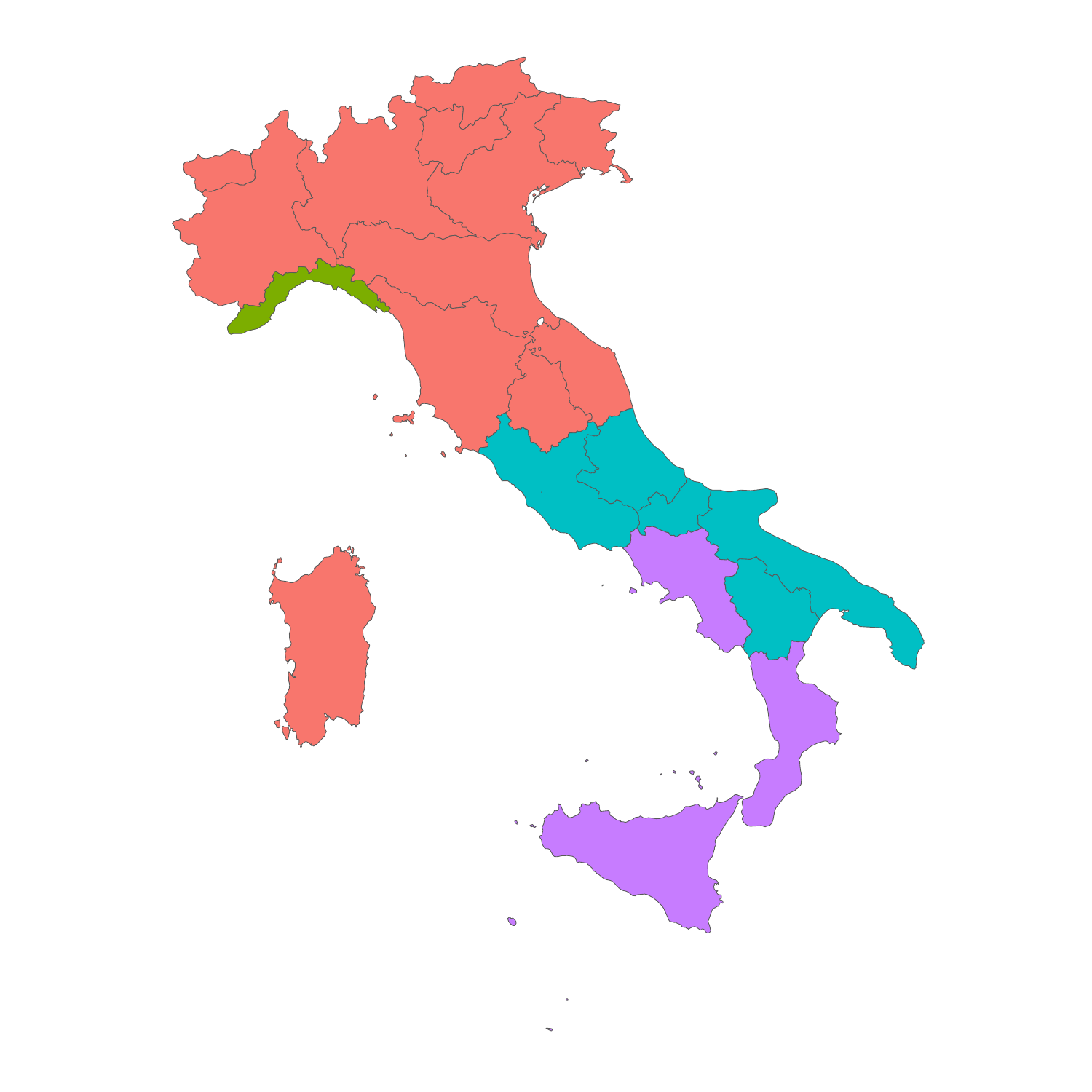}
         \caption{clusters}
         \label{subfig:italian_regions}
     \end{subfigure}
     \hspace{0.1em}
         \begin{subfigure}[b]{0.47\textwidth}
         \centering
         \includegraphics[width=\textwidth]{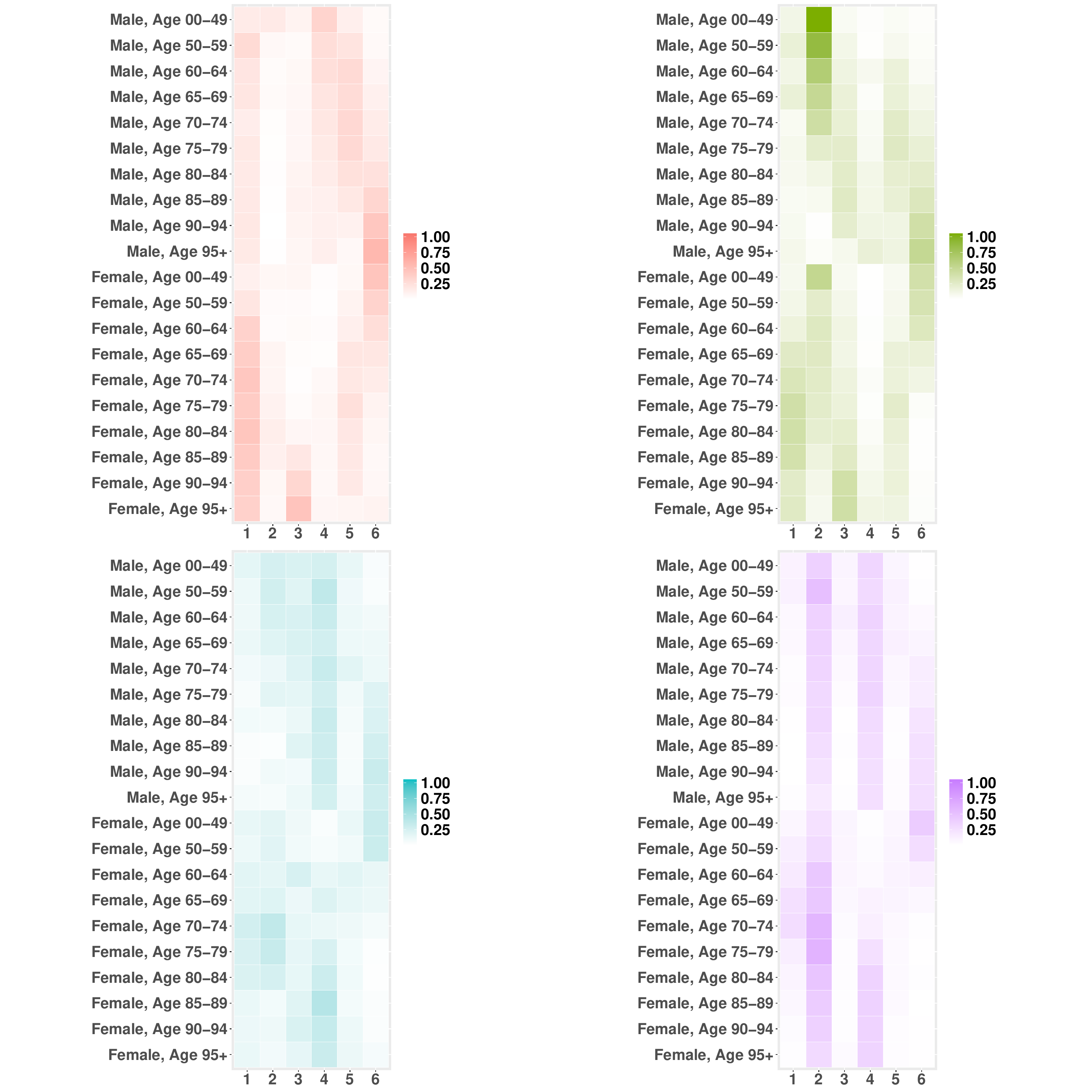}
         \caption{medoids}
         \label{subfig:medoids}
     \end{subfigure}
     \caption{PAM classification of Italian regions based on $\lambda^{(1)}_{i,h_1}$. Horizontal axces in \subref{subfig:medoids} denote latent classes.}
     \label{fig:lambda1}
\end{figure}

Three blocks of latent parameters are introduced in the BPRTTD model and they are arranged in a dependent structure; that is, each latent class $\lambda^{(1)}_{i,h_1}, h_1=1,\dots,H_1$ is characterized by different 
$\lambda^{(2)}_{t,h_1,h_2}$, and furthermore $h_2$-specific $\lambda^{(3)}_{k,h_2}, h_2=1,\dots,H_2$. Therefore we approach the interpretation of latent parameters in an orderly manner. We start with the first block of latent parameters $\lambda^{(1)}_{i,h_1}$ that allocate demographic groups defined by Italian regions, gender and age groups into $H_1$ latent classes. Table \ref{tb:lambda1_real} in the \nameref{Appendices} shows the posterior mean estimates of $\lambda^{(1)}_{i,h_1}$ and we highlight in red values above the mean $\alpha_1/\alpha_2$ of the Gamma prior distribution. Note that since we already have a Poisson regression component that accounts for global linear relationships between covariates and death rates, what we see in estimates of $\lambda^{(1)}_{i,h_1}$ indicates differential local effects of higher order interactions between covariates on mortality rates unexplained by linear regression. It is clear that although latent classes labeled by $h_1=1$ and $h_1=4$ represent majorly female and male mortality patterns respectively, they appear to be geographical dependent. For instance, almost all female age groups, except for older female (age group 85+) from southern Italy (Molise, Campania, Apulia, Basilicata, Calabria, Sicily) show elevated weights in latent class $h_1=1$ in Table \ref{tb:lambda1_real1}, whereas the same older female southern Italian population shares similar mortality patterns with almost all male groups, excluding those in northern Italy (Piemonte, Valle d’Aosta, Lombardia, Veneto, Friuli-Venezia Giulia, Emilia-Romagna) as shown in Table \ref{tb:lambda1_real4}. Latent class $h_1=6$, on the other hand, suggests old male and young female share something in common in their causes of death over time captured by $\lambda^{(2)}_{t,6,h_2}$ and $\lambda^{(3)}_{k,h_2}$. The remaining three latent classes indexed by $h_1=2,3,5$ are less related to age and gender but show more geographical dependence. So before we move on to the analysis of latent parameters $\lambda^{(2)}_{t,h_1,h_2}$ and $\lambda^{(3)}_{k,h_2}$, we make another attempt to decipher the local joint effect of regions, age and gender. To do this, we first rearrange the posterior mean estimates of $\lambda^{(1)}_{i,h_1}, i=1,\dots,N, h_1=1,\dots,H_1$ into a new matrix of dimension $21\times (2\times 10\times H_1)$ where 21 is the number of Italian regions, 2 and 10 are gender and age groups. Then we treat 21 Italian regions as observations, gender, age groups as well as $H_1$ latent classes as features, apply the partitioning around medoids (PAM) algorithm to classify Italian regions based on features. Optimal number of clusters is 4 according to the elbow method. The clustering algorithm confirms previous observations. Figure \ref{subfig:italian_regions} shows that northern Italy plus Toscana, Umbria, Marche is classified in a different group from southern Italy, plus Lazio and excluding Campania, Calabria, Sicily. Although the two clusters have similar weights in latent class $h_1=6$, the differences mainly exist in latent class $h_1=1$ for female population in the northern Italy cluster and $h_1=4$ for older female in the southern Italy cluster. The PAM algorithm also separate the conventional classification of southern Italy further into two groups that exhibit homogeneous behavior when looking at latent class $h_1=4$ and $h_1=6$, but differ quite substantially in latent class $h_2$, see Figure \ref{subfig:medoids}. Lastly, Liguria is singled out to form its own cluster because latent class $h_1=2$ plays a rather significant role in defining the mortality pattern over time in the region.

\begin{figure}
     \centering
    \begin{subfigure}[b]{0.31\textwidth}
         \centering
         \includegraphics[width=\textwidth]{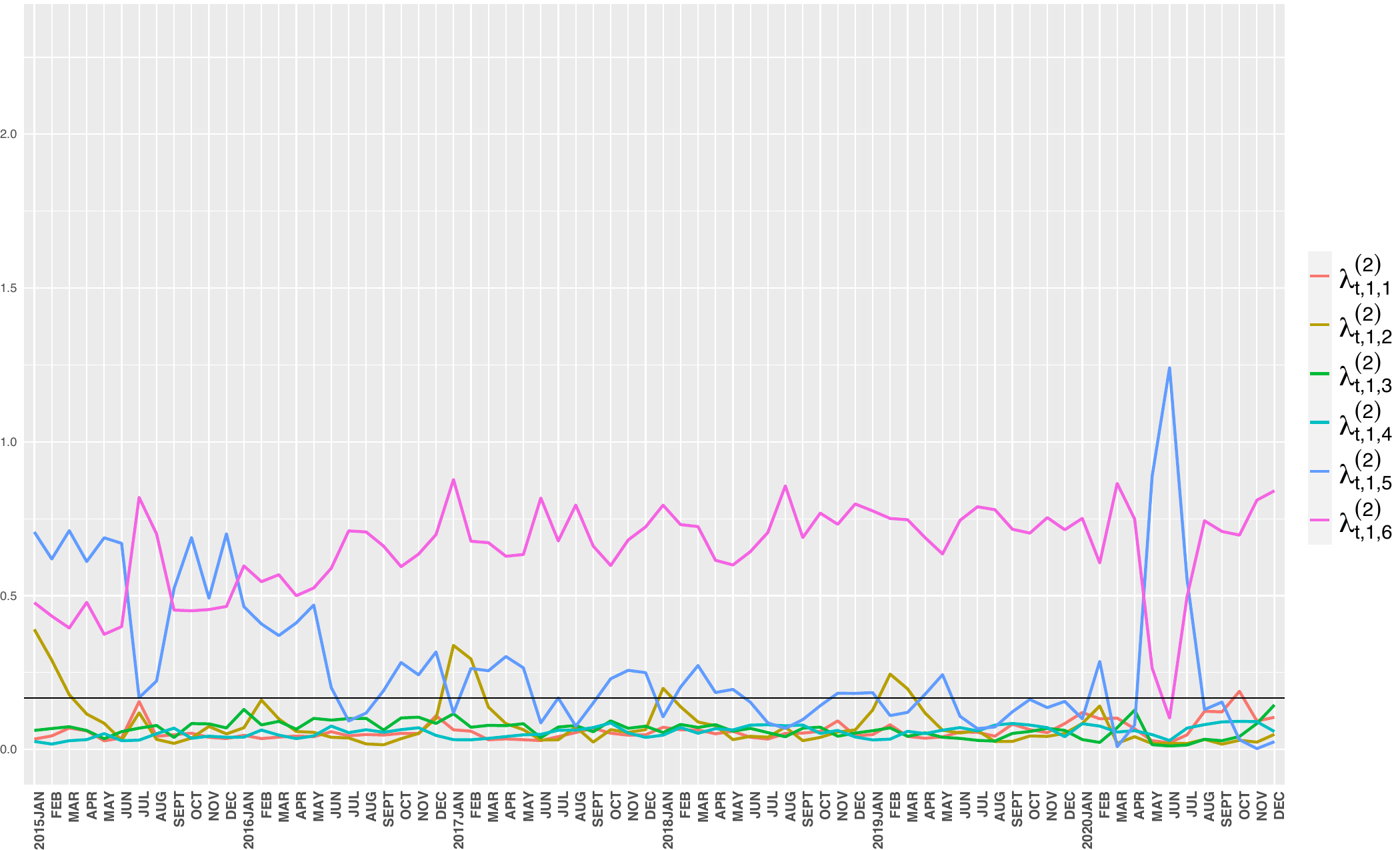}
         \caption{$\lambda^{(2)}_{t,1,h_2}$}
         \label{subfigure:lambda2_1}
     \end{subfigure}
     \hspace{0.1em}
         \begin{subfigure}[b]{0.31\textwidth}
         \centering
         \includegraphics[width=\textwidth]{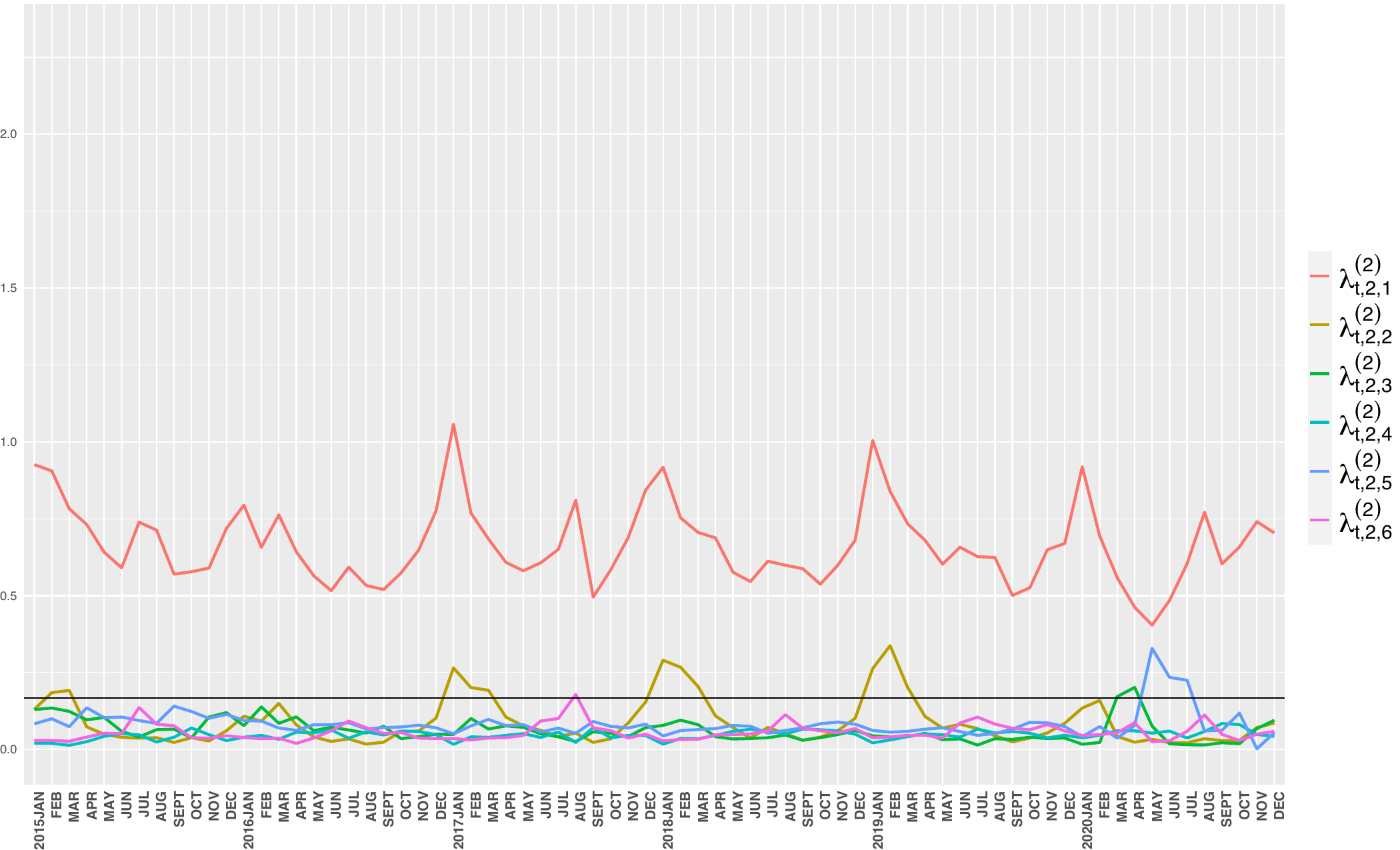}
         \caption{$\lambda^{(2)}_{t,2,h_2}$}
         \label{subfigure:lambda2_2}
     \end{subfigure}
       \hspace{0.1em}
         \begin{subfigure}[b]{0.31\textwidth}
         \centering
         \includegraphics[width=\textwidth]{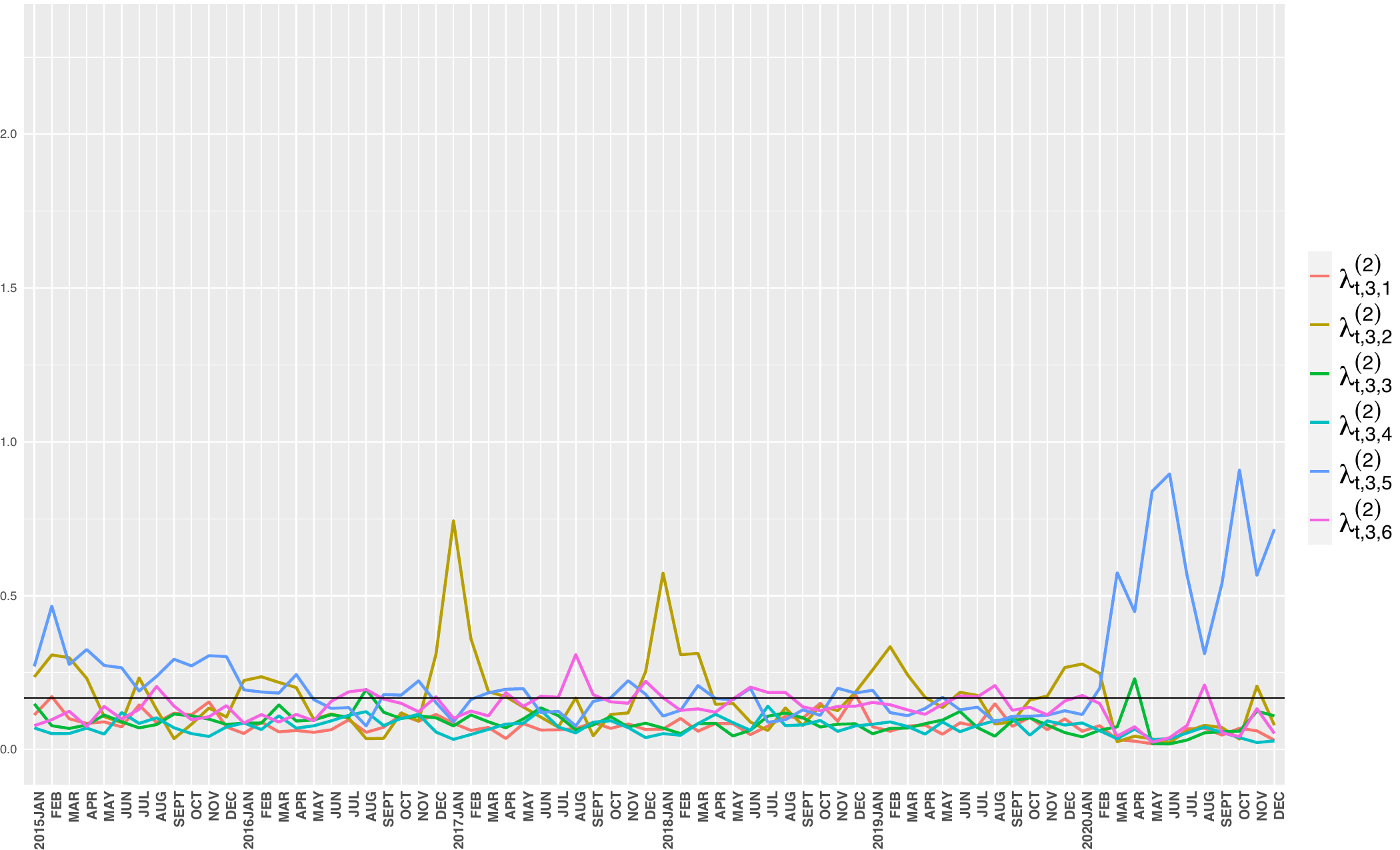}
         \caption{$\lambda^{(2)}_{t,3,h_2}$}
         \label{subfigure:lambda2_3}
     \end{subfigure}
         \begin{subfigure}[b]{0.31\textwidth}
         \centering
         \includegraphics[width=\textwidth]{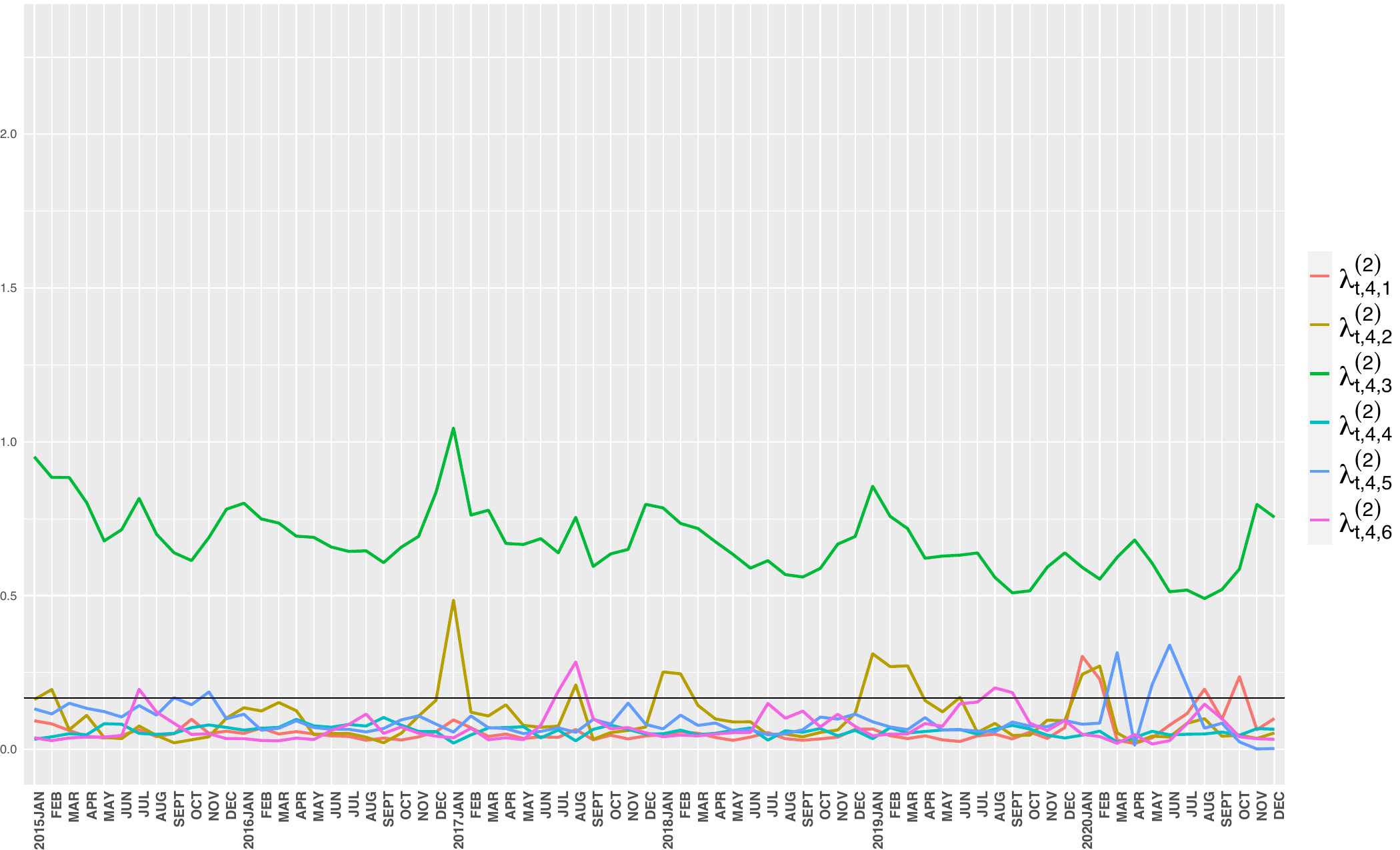}
         \caption{$\lambda^{(2)}_{t,4,h_2}$}
         \label{subfigure:lambda2_4}
     \end{subfigure}
             \hspace{0.1em}
         \begin{subfigure}[b]{0.31\textwidth}
         \centering
         \includegraphics[width=\textwidth]{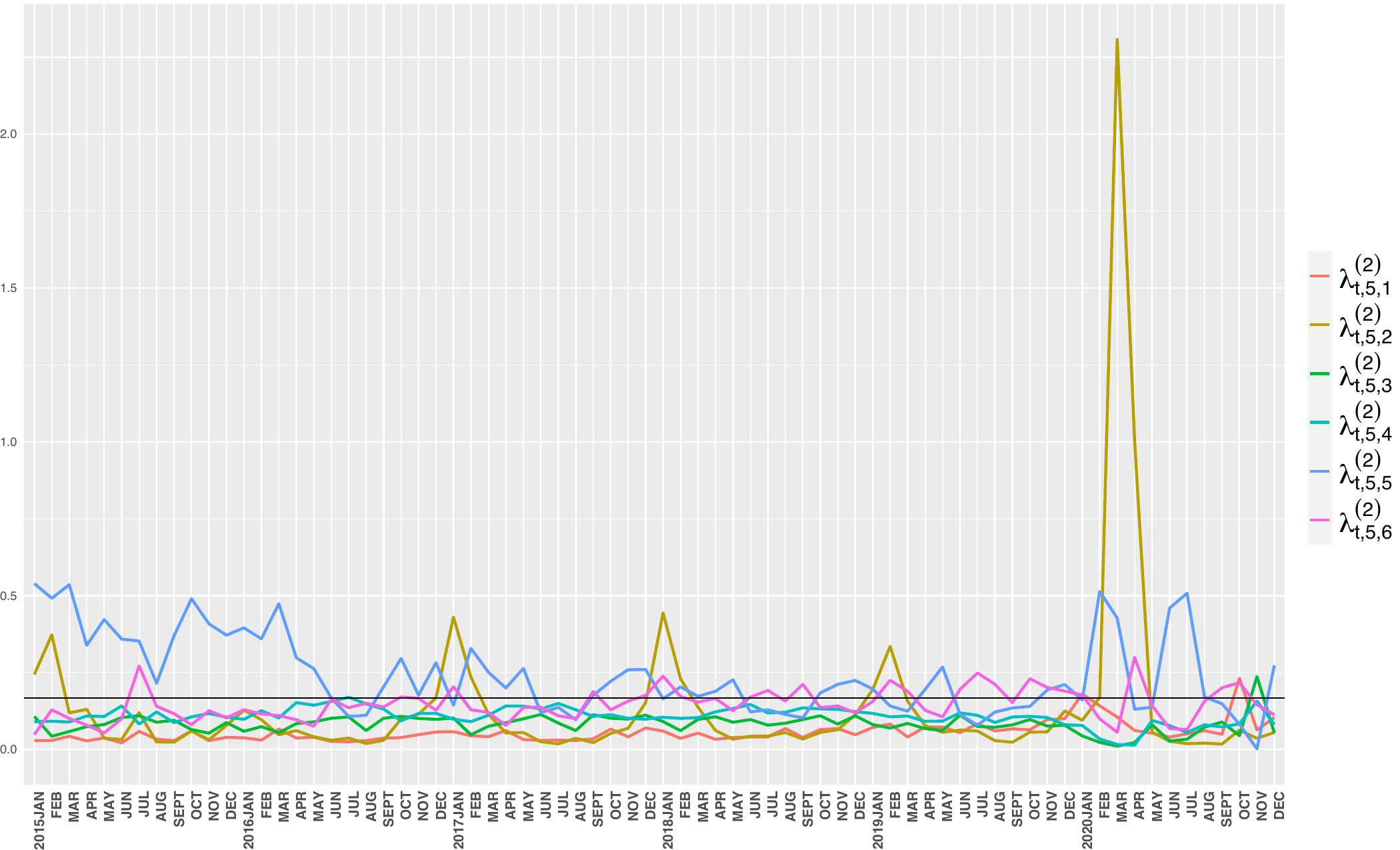}
         \caption{$\lambda^{(2)}_{t,5,h_2}$}
         \label{subfigure:lambda2_5}
     \end{subfigure}
             \hspace{0.1em}
         \begin{subfigure}[b]{0.31\textwidth}
         \centering
         \includegraphics[width=\textwidth]{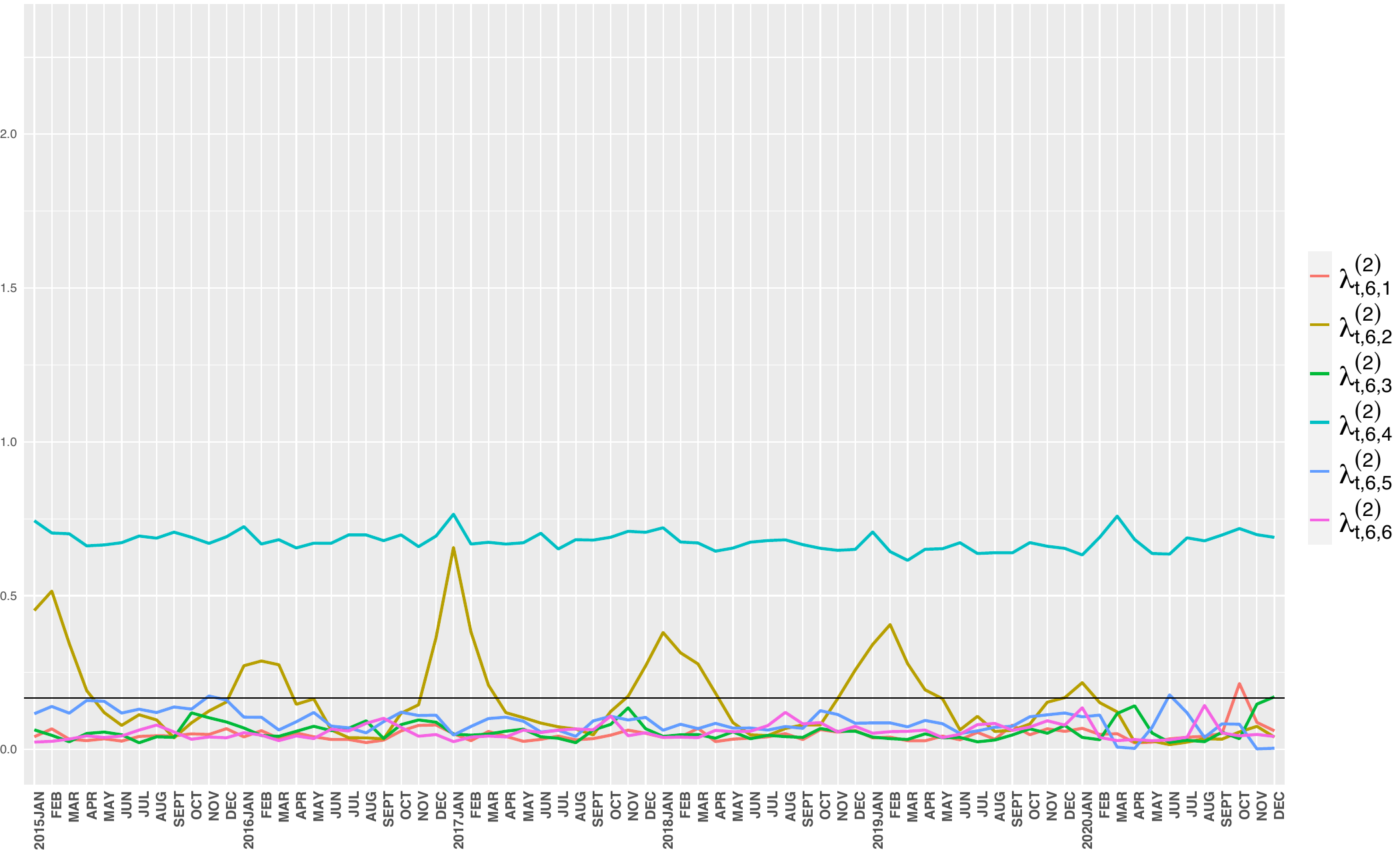}
         \caption{$\lambda^{(2)}_{t,6,h_2}$}
         \label{subfigure:lambda2_6}
     \end{subfigure}
     \caption{Trajectories of $\lambda^{(2)}_{t,h_1,h_2}$ for each latent $h_1=1,\dots,H_1$ from January 2015 to December 2020. Black horizontal lines stand for the Gamma prior mean $\beta_1/\beta_2$ on $\lambda^{(2)}_{t,h_1,h_2}$.}
     \label{fig:lambda2}
\end{figure}

\begin{figure}
     \centering
    \begin{subfigure}[b]{0.31\textwidth}
         \centering
         \includegraphics[width=\textwidth]{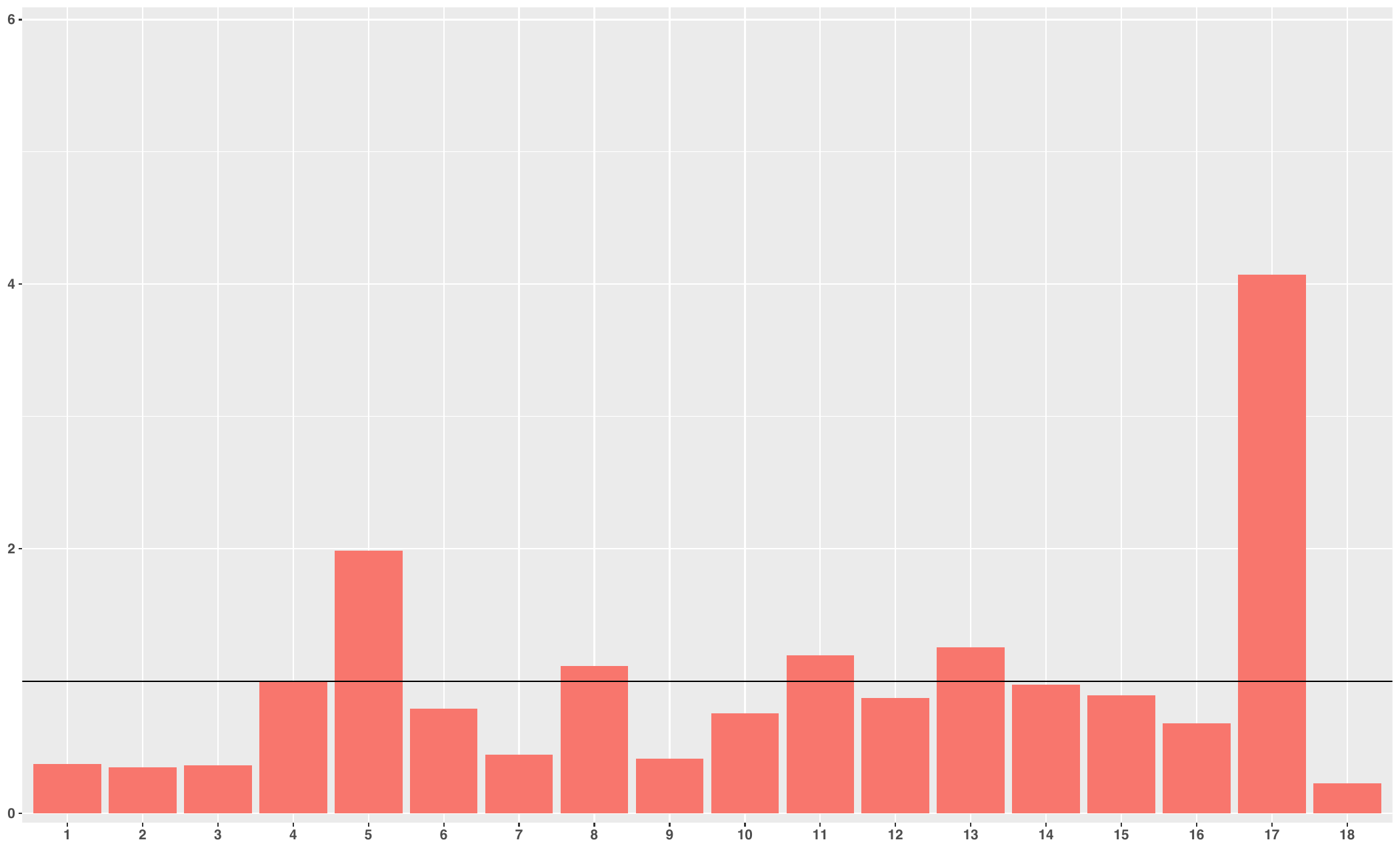}
         \caption{$\lambda^{(3)}_{k,1}$}
         \label{subfigure:lambda3_1}
     \end{subfigure}
     \hspace{0.1em}
         \begin{subfigure}[b]{0.31\textwidth}
         \centering
         \includegraphics[width=\textwidth]{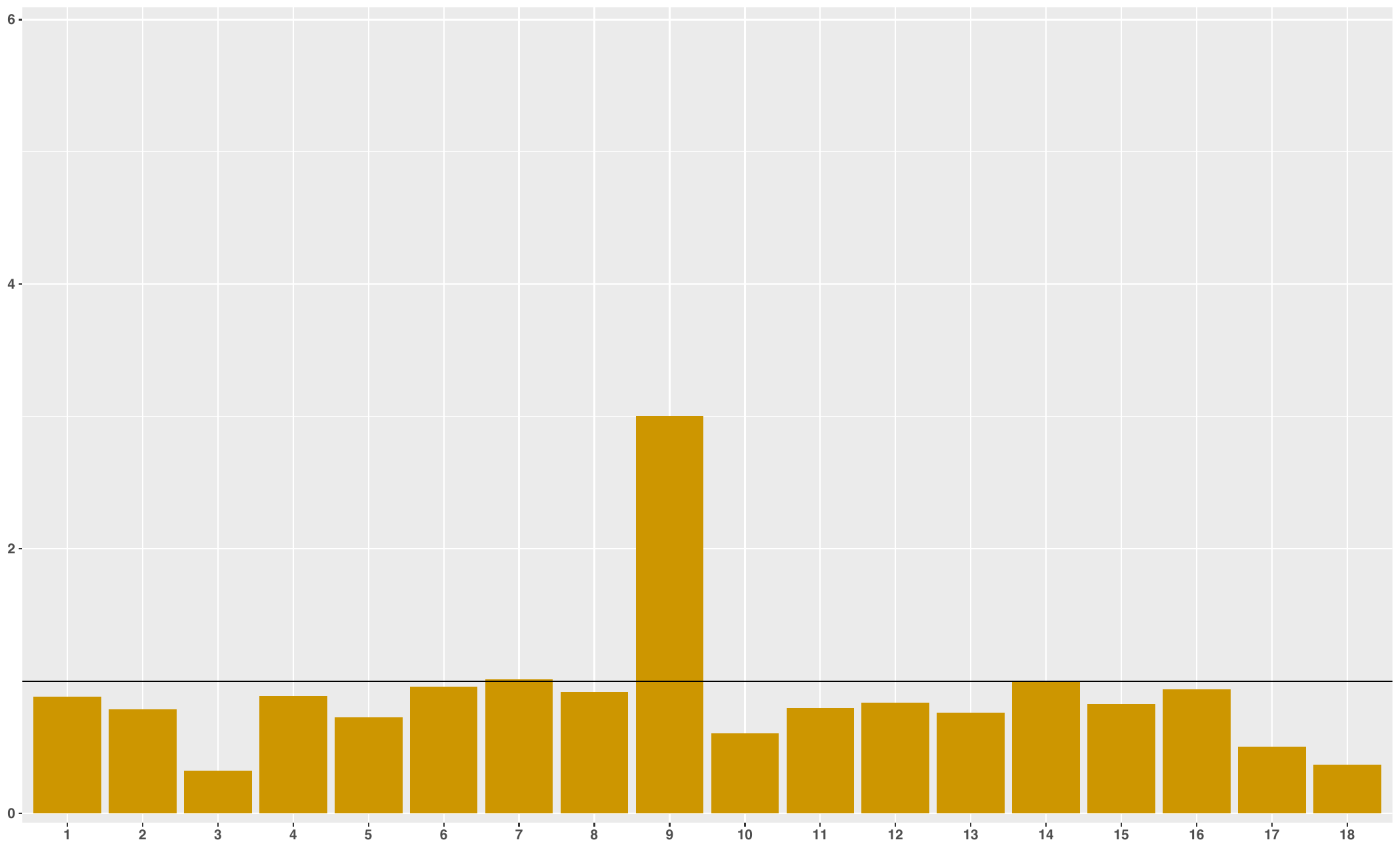}
         \caption{$\lambda^{(3)}_{k,2}$}
         \label{subfigure:lambda3_2}
     \end{subfigure}
       \hspace{0.1em}
         \begin{subfigure}[b]{0.31\textwidth}
         \centering
         \includegraphics[width=\textwidth]{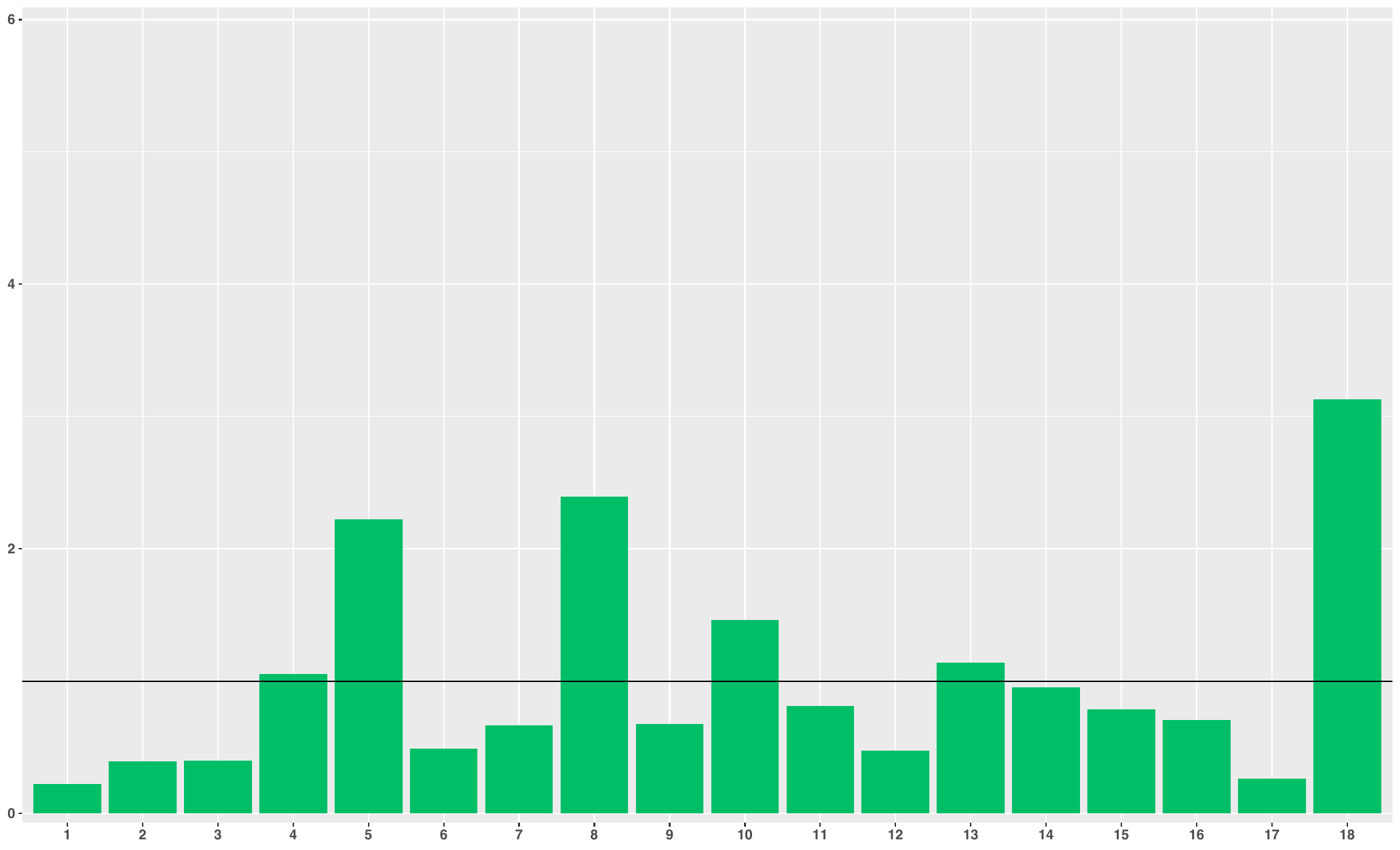}
         \caption{$\lambda^{(3)}_{k,3}$}
         \label{subfigure:lambda3_3}
     \end{subfigure}
         \begin{subfigure}[b]{0.31\textwidth}
         \centering
         \includegraphics[width=\textwidth]{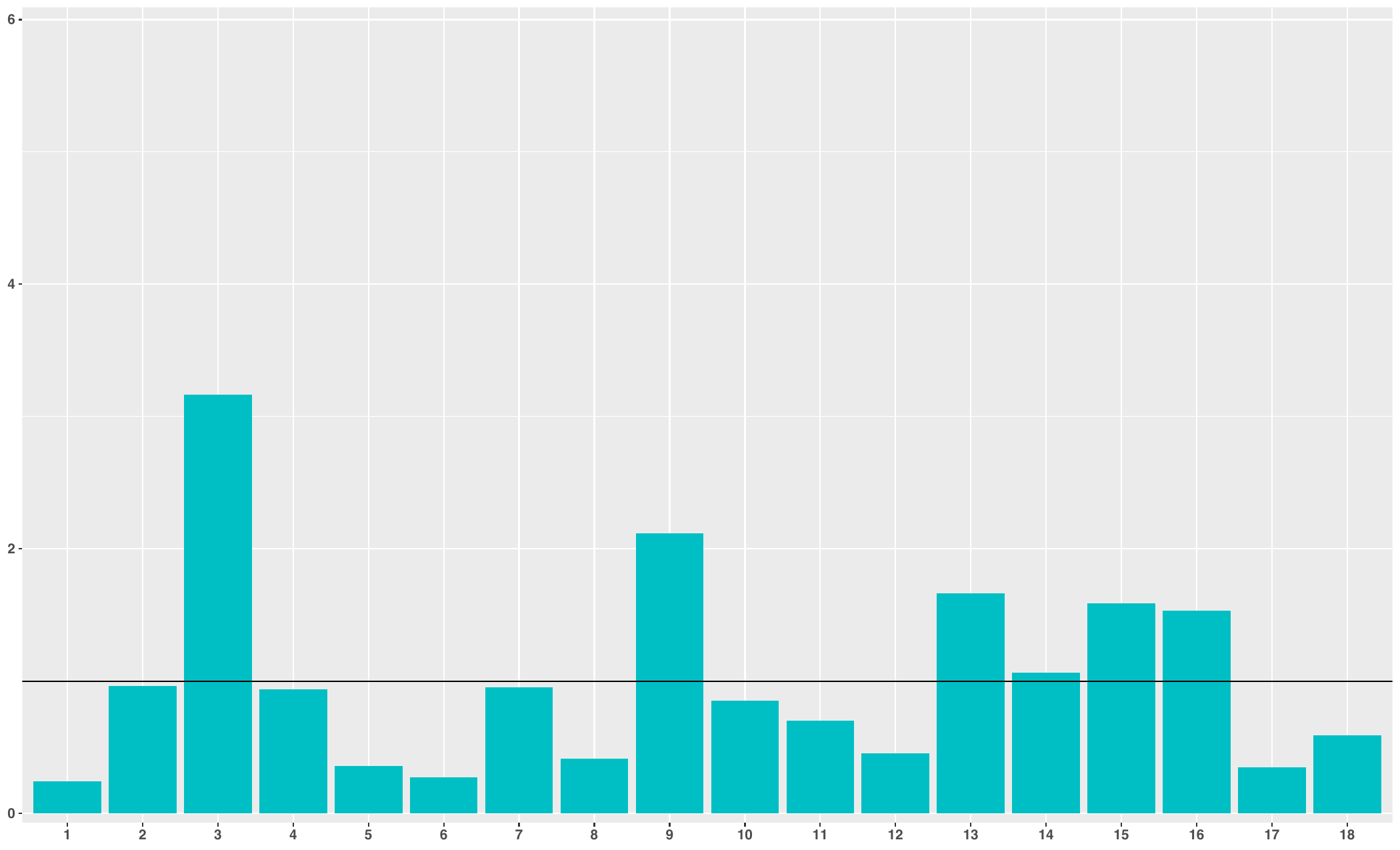}
         \caption{$\lambda^{(3)}_{k,4}$}
         \label{subfigure:lambda3_4}
     \end{subfigure}
             \hspace{0.1em}
         \begin{subfigure}[b]{0.31\textwidth}
         \centering
         \includegraphics[width=\textwidth]{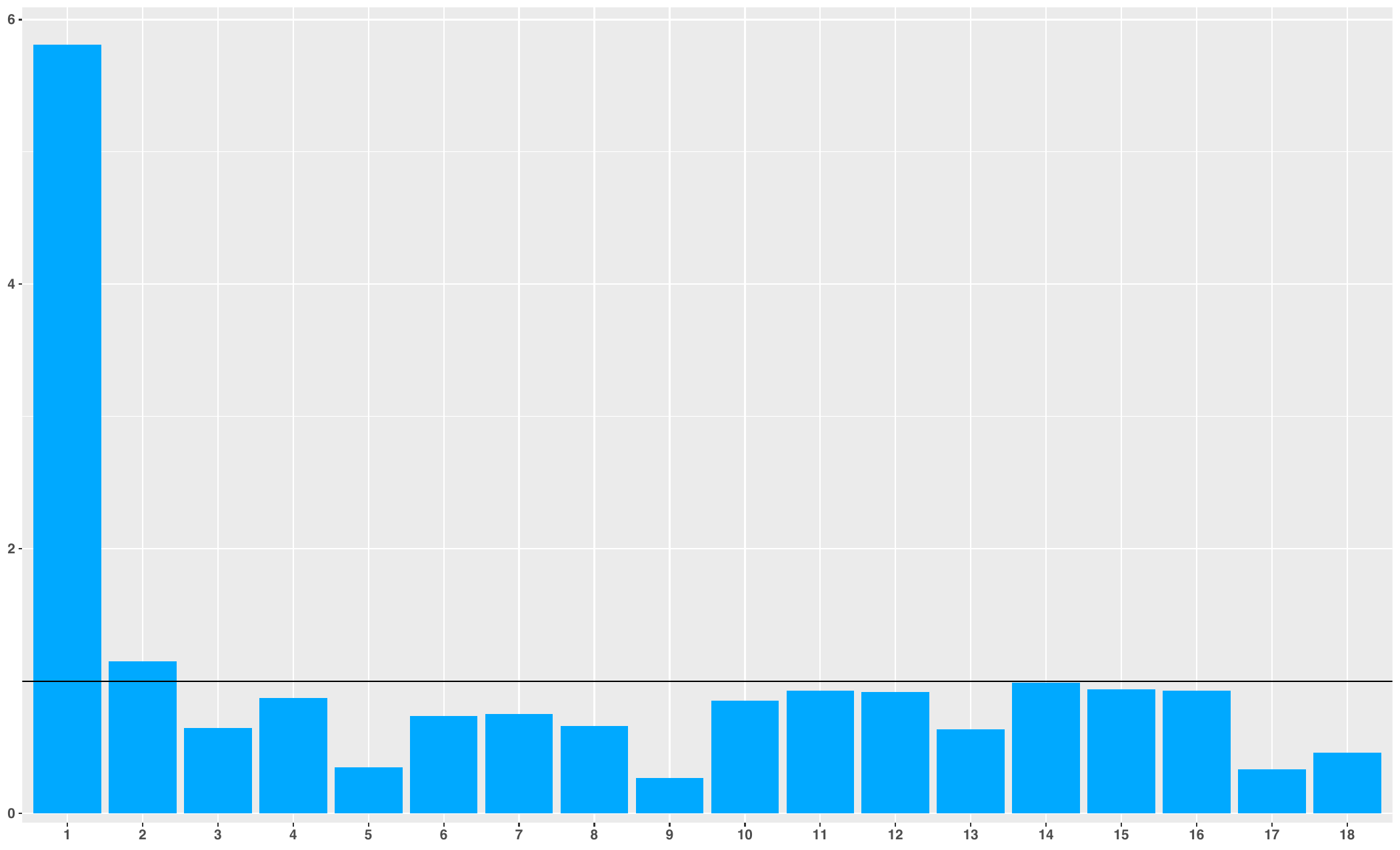}
         \caption{$\lambda^{(3)}_{k,5}$}
         \label{subfigure:lambda3_5}
     \end{subfigure}
             \hspace{0.1em}
         \begin{subfigure}[b]{0.31\textwidth}
         \centering
         \includegraphics[width=\textwidth]{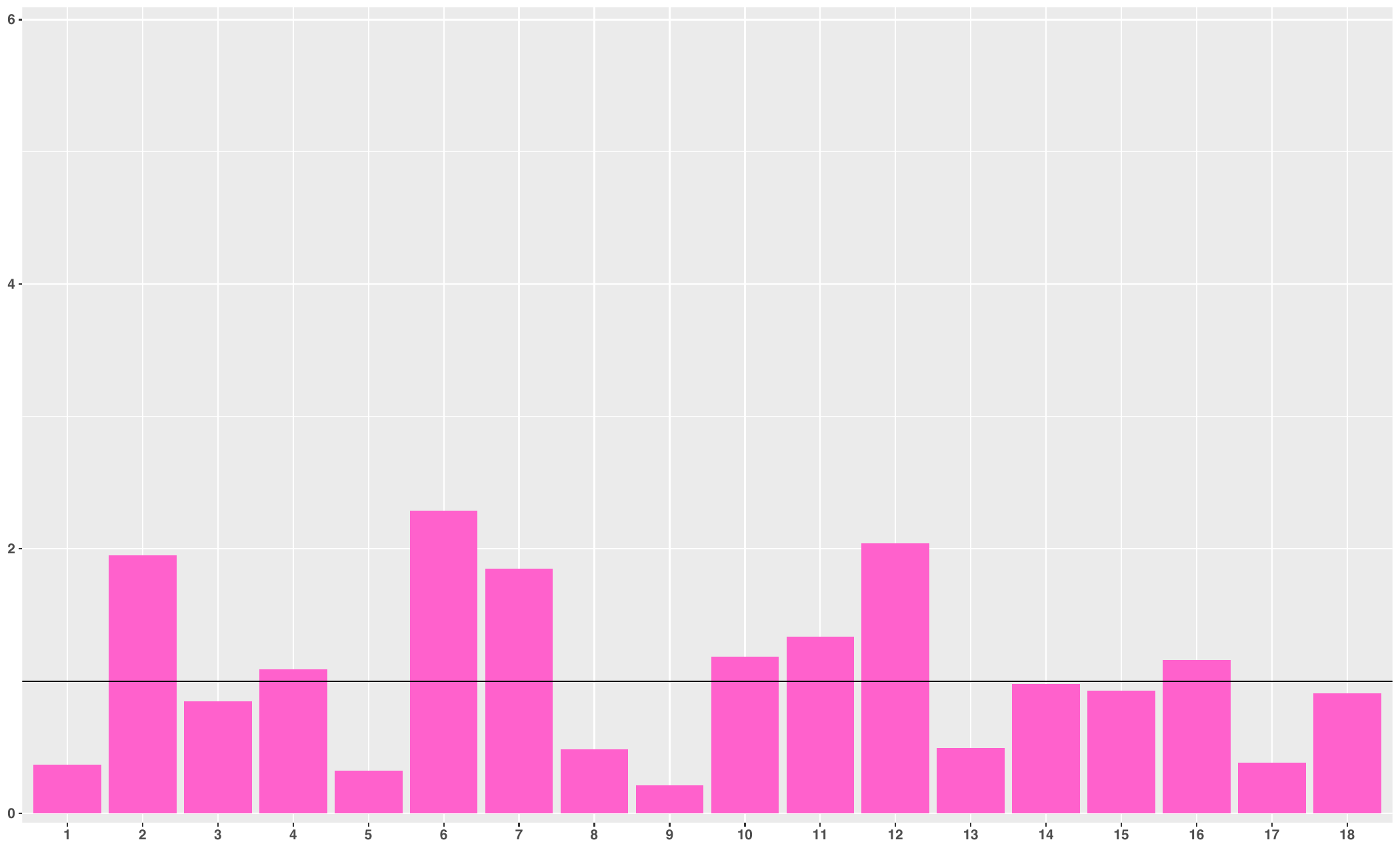}
         \caption{$\lambda^{(3)}_{k,6}$}
         \label{subfigure:lambda3_6}
     \end{subfigure}
     \caption{Bar plots of $\lambda^{(3)}_{k,h_2}$ of 18 causes of death (horizontal axes) for each latent $h_2=1,\dots,H_2$. Black horizontal lines stand for the Gamma prior mean $\epsilon_1/\epsilon_2$ on $\lambda^{(3)}_{k,h_2}$.}
     \label{fig:lambda3}
\end{figure}

We proceed to analyze together the second layer latent parameters $\lambda^{(2)}_{t,h_1,h_2}$ and the third layer latent parameters $\lambda^{(3)}_{k,h_2}$ as they jointly identify the corresponding latent classes labeled by $h_1$. $\lambda^{(2)}_{t,h_1,h_2}$ is the block of parameters associated with time indices $T$, so we display the posterior mean estimates of $\lambda^{(2)}_{t,h_1,h_2}$ in terms of trajectories evolving over time in Figure \ref{fig:lambda2}; on the other hand, $\lambda^{(3)}_{k,h_2}$ utilizes $H_2$ latent structures to summarizes 18 causes of death as shown in Figure \ref{fig:lambda3}. We begin with latent class $h_1=1$ shown in Figure \ref{subfigure:lambda2_1} that is significant for almost all female age groups except for older population in southern Italy. Two trajectories are more relevant in this class, and they are characterized by mortality rates $\lambda^{(3)}_{k,5}$ in Figure \ref{subfigure:lambda3_5} and $\lambda^{(3)}_{k,6}$ in Figure \ref{subfigure:lambda3_6}. $\lambda^{(3)}_{k,5}$ mostly captures COVID-19 mortality and the trajectory $\lambda_{t,1,5}^{(1)}$ in Figure \ref{subfigure:lambda2_1} indicates a sudden weight spike of this particular latent class $h_2=5$ around June 2020. This is when the pandemic situation eases between the first wave and the second wave so the daily new cases are almost single digits; the time lag between contracting COVID-19 in the previous wave and dying of COVID-19 potentially results in the spike that we observe. We will see later another type of weight spike with respect to latent class $h_2=5$. $\lambda_{t,1,5}^{(1)}$ is also active from January 2015 to July 2016. However, since during this period the new cases offset in the BPRTTD model is exactly 0, the dominating factor is no longer COVID-19 death, but possibly the other cause in $\lambda^{(3)}_{k,5}$ higher than the prior mean, namely some infectious and parasitic diseases. Trajectory $\lambda_{t,1,6}^{(1)}$ has opposite behavior as $\lambda_{t,1,5}^{(1)}$; that is, it is squeezed out when the latter is high and bounces when the latter is low. When we inspect $\lambda_{k,6}^{(3)}$ in Figure \ref{subfigure:lambda3_6}, the following causes of death have rising weights including 2. Some infectious and parasitic diseases, 6. Psychic and behavioral disorders, 7. Diseases of the nervous system and sense organs, 10. Diseases of the digestive system, 11. Diseases of the skin and subcutaneous tissue and 12. Diseases of the musculoskeletal system and connective tissue. In the Poisson regression component, we observe positive global main effect of COVID lockdown measures on the mortality rate of psychic and behavioral disorders, however, the squeezing phenomenon does not contradict our previous arguments; in fact, since we are discussing latent class $h_1=1$ crucial to female population except for older ones in southern Italy, it actually suggests a local compensation effect specific to this demographic group.

We have commented beforehand that latent class $h_1=2$ are unique to three southern Italian regions, Campania, Calabria and Sicily and now we see that the determining trajectory $\lambda_{t,2,1}^{(2)}$ in Figure \ref{subfigure:lambda2_2} has high estimated rates in causes of death 5. Endocrine, nutritional and metabolic diseases as well as 17. Symptoms, signs, abnormal results and ill-defined causes in Figure \ref{subfigure:lambda3_1}. It shows strong seasonality with peaks in both winter and summer. Endocrine, nutritional and metabolic diseases have been documented to be related to winter holidays \citep{phillips2010christmas} and heat exposure \citep{zhao2019assessment}. On the other hand, the seasonality of symptoms, signs, abnormal results and ill-defined causes in these three regions may consist of misclassified deaths related to seasonal illnesses. Latent class $h_1=3$ is almost exclusively explanatory for female older than 85 years old in northern Italy and some male age groups in the south. The class portraits a pattern where COVID-19 mortality rate goes through two spikes in June 2020 and October 2020 in Figure \ref{subfigure:lambda3_3}. As stated in the previous paragraph, the spike in the end of the first wave is possibly due to the lag between contracting and death of COVID-19; the spike in mortality rate in October anticipates the strike of second COVID-19 wave. This can be the outcome of many factors, for instance, even though Italy has gone through the first wave, in face of second wave, testing and reporting of COVID-19 cases are still insufficient, leading to an underestimating of real case number. The health system is also not thoroughly prepared to combat the much intenser comeback of COVID-19 in the coming fall and winter. We distinguish two types of displacements between case peak and mortality peak. The first type is usually seen after a previous COVID-19 wave and it is due to the time lag between contracting COVID and final death whereas the second type predicts the incoming COVID-19 hit, which is particular true in 2020 when the society and the health system are seriously under prepared to tackle the pandemic. Additionally recall that this represents local effects for female older than 85 years old in northern Italy and certain male age groups in the south, suggesting that underpreparedness is particular detrimental to those people. We also notice that the trajectories of all other causes of death are crowded out by $\lambda_{t,3,5}^{(2)}$ in 2020, offering evidences to the hypothesis that potential harvesting effect exists. Next latent class $h_1=4$ underlies the mortality composition of young male Italian population in the north, all male and female population in the south. The essential feature of this class is the downward trend of trajectory $\lambda_{t,4,3}^{(2)}$ displayed in Figure \ref{subfigure:lambda2_4}. A closer look at Figure \ref{subfigure:lambda3_3} reveals that 5. Endocrine, nutritional and metabolic diseases, 8. Diseases of the circulatory system and 18. External causes of trauma and poisoning are the three causes that define the mortality structure in $\lambda^{(3)}_{k,3}$. The trajectories indicates these three mortality causes tend to be seasonal. Although we have briefly commented on the seasonality of mortality due to endocrine, nutritional and metabolic diseases observed in Campania, Calabria and Sicily, we elaborate on the fact that the seasonality is distinct with young male Italian population in the north, all male and female population in the south except for the three regions just mentioned. The spikes are generally less drastic in the second demographic group. For instance, when heatwave hits Campania, Calabria and Sicily in the summer of 2017, causing noticeable increase in the number of people dying of endocrine, nutritional and metabolic diseases, the situation is less severe in the north. Another observation worth pointing out is that endocrine, nutritional and metabolic diseases are more lethal for older female population as indicated in Table \ref{tb:lambda1_real2} and Table \ref{tb:lambda1_real4} in the \nameref{Appendices}. Diseases of the circulatory system are causes of death whose seasonality has been widely studies as well and our findings concur with previous findings in the literature \citep{fares2013winter, stewart2017seasonal}. Lastly, the seasonality of external causes of trauma and poisoning may largely be contributed to increasing traffic accidents in the winter and outdoor activities in the summer.

Latent class $h_1=5$ in Figure \ref{subfigure:lambda2_5}, which is primarily significant for both male and female in northern Italy, has two major attributes. One is that trajectory $\lambda_{t,5,2}^{(2)}$ representing mortality rate of 9. Diseases of the respiratory system shows an abnormal spike around March and April 2020 when health system is overwhelmed in northern Italy and many COVID-19 deaths are mis-classified. Similar argument has been made when we interpret the coefficients of Poisson regression component of the BPRTTD model. The other attribute that characterizes the latent class is $\lambda_{t,5,5}^{(2)}$ with its two peaks first in February and and then July 2020. Both types of displacement of COVID mortality rate appear. Almost all male between the age 50 and 89 and female between 70 and 94 in northern Italy except for Veneto and Friuli-Venezia Giulia experience the second type of displacement and are subjects to elevated mortality rate of dying of COVID-19 in the beginning of first wave (February and March 2020). On the contrary, the second type occurs to older female population in northern Italy and certain male age groups in the south only in the beginning of second wave as previously illustrated. We close the analysis by commenting on latent class $h_1=6$ shown in Figure \ref{subfigure:lambda2_6} which features constant trend of $\lambda^{(2)}_{t,6,4}$ defined mostly by tumor and respiratory diseases shown in Figure \ref{subfigure:lambda3_4}. Another relevant trajectory $\lambda^{(2)}_{t,6,2}$ captures expected seasonality of respiratory disease deaths. This is the mortality structure shared by male older population and female population under 69 across almost all Italian regions.

\section{Summary and Future Work}
\label{Summary and Future Work}

In this paper, we propose to model Poisson count data using the BPRTTD model. The model comprises two parts, the first part is the Poisson regression model. In the second part, the data are organized as tensor and we apply tensor train decomposition to estimate latent parameter space for explanatory purposes. The Bayesian inference framework is validated in two simulation studies and then applied to the Italian monthly causes specific mortality data from January 2015 to December 2020. The regression component leverages information in covariates and we are able to identify causes of death that are positively, negatively and not related to government interventions during the COVID-19 pandemic. We also discover the joint effects of age, gender and causes of death on mortality rate via the tensor decomposition component that compensates what the Poisson regression fails to account for. It enables a further stratification of demographic profiles characterized jointly by geographical location, gender and age based on their unique dynamic mortality structures over the time span. Regional classification are made and the results coincide with conventional conception. COVID-19 related consequences are also revealed in the latent parameters. Several causes of death, including infectious and parasitic diseases and psychic and behavioral disorders,  compete with COVID-19 mortality among specific demographic groups. 

In the BPRTTD model, we have not fully exploit the spatial-temporal information in the data. For instance, instead of applying clustering algorithms to the posterior estimates, one can introduce reasonable distance measures and utilize geographic locations encoded in $\lambda_{i,h_1}^{(1)}$ when specifying the model. $\lambda_{t,h_1,h_2}^{(2}$ can also be modeled in a time series framework so that the temporal dependence can be inferred. Another possible future direction is the proper choice of tensor train ranks in the BPRTTD model plays an important role in controlling the model complexity. The model selection can be accomplished by calculating marginal likelihoods over a pre-specified grids defined by tensor train ranks. Due to the increased computational burden this solution would require, we leave its exploration to future work.

\newpage

\bibliographystyle{apalike}
\bibliography{ref}

\begin{thebibliography}{}

\bibitem[Abedi et~al., 2021]{abedi2021racial}
Abedi, V., Olulana, O., Avula, V., Chaudhary, D., Khan, A., Shahjouei, S., Li,
  J., and Zand, R. (2021).
\newblock Racial, economic, and health inequality and covid-19 infection in the
  united states.
\newblock {\em Journal of racial and ethnic health disparities}, 8:732--742.

\bibitem[Aveyard et~al., 2021]{aveyard2021association}
Aveyard, P., Gao, M., Lindson, N., Hartmann-Boyce, J., Watkinson, P., Young,
  D., Coupland, C.~A., San~Tan, P., Clift, A.~K., Harrison, D., et~al. (2021).
\newblock Association between pre-existing respiratory disease and its
  treatment, and severe covid-19: a population cohort study.
\newblock {\em The Lancet Respiratory Medicine}, 9(8):909--923.

\bibitem[Bambra et~al., 2020]{bambra2020covid}
Bambra, C., Riordan, R., Ford, J., and Matthews, F. (2020).
\newblock The covid-19 pandemic and health inequalities.
\newblock {\em J Epidemiol Community Health}, 74(11):964--968.

\bibitem[Bol et~al., 2021]{bol2021effect}
Bol, D., Giani, M., Blais, A., and Loewen, P.~J. (2021).
\newblock The effect of covid-19 lockdowns on political support: Some good news
  for democracy?
\newblock {\em European journal of political research}, 60(2):497--505.

\bibitem[Britton et~al., 2020]{britton2020covid}
Britton, P.~N., Hu, N., Saravanos, G., Shrapnel, J., Davis, J., Snelling, T.,
  Dalby-Payne, J., Kesson, A.~M., Wood, N., Macartney, K., et~al. (2020).
\newblock Covid-19 public health measures and respiratory syncytial virus.
\newblock {\em The Lancet Child \& Adolescent Health}, 4(11):e42--e43.

\bibitem[Cai et~al., 2022]{cai2022generalized}
Cai, J.-F., Li, J., and Xia, D. (2022).
\newblock Generalized low-rank plus sparse tensor estimation by fast riemannian
  optimization.
\newblock {\em Journal of the American Statistical Association}, pages 1--17.

\bibitem[Calderon-Anyosa and Kaufman, 2021]{calderon2021impact}
Calderon-Anyosa, R.~J. and Kaufman, J.~S. (2021).
\newblock Impact of covid-19 lockdown policy on homicide, suicide, and motor
  vehicle deaths in peru.
\newblock {\em Preventive Medicine}, 143:106331.

\bibitem[Cheval et~al., 2020]{cheval2020observed}
Cheval, S., Mihai~Adamescu, C., Georgiadis, T., Herrnegger, M., Piticar, A.,
  and Legates, D.~R. (2020).
\newblock Observed and potential impacts of the covid-19 pandemic on the
  environment.
\newblock {\em International journal of environmental research and public
  health}, 17(11):4140.

\bibitem[Cichocki et~al., 2016]{cichocki2016tensor}
Cichocki, A., Lee, N., Oseledets, I., Phan, A.-H., Zhao, Q., Mandic, D.~P.,
  et~al. (2016).
\newblock Tensor networks for dimensionality reduction and large-scale
  optimization: Part 1 low-rank tensor decompositions.
\newblock {\em Foundations and Trends{\textregistered} in Machine Learning},
  9(4-5):249--429.

\bibitem[Conteduca and Borin, 2022]{conteduca2022new}
Conteduca, F.~P. and Borin, A. (2022).
\newblock A new dataset for local and national covid-19-related restrictions in
  italy.
\newblock {\em Italian Economic Journal}, 8(2):435--470.

\bibitem[Dadras et~al., 2021]{dadras2021effects}
Dadras, O., Alinaghi, S. A.~S., Karimi, A., MohsseniPour, M., Barzegary, A.,
  Vahedi, F., Pashaei, Z., Mirzapour, P., Fakhfouri, A., Zargari, G., et~al.
  (2021).
\newblock Effects of covid-19 prevention procedures on other common infections:
  a systematic review.
\newblock {\em European Journal of Medical Research}, 26(1):1--13.

\bibitem[Dmetrichuk et~al., 2022]{dmetrichuk2022retrospective}
Dmetrichuk, J.~M., Rosenthal, J.~S., Man, J., Cullip, M., and Wells, R.~A.
  (2022).
\newblock Retrospective study of non-natural manners of death in ontario:
  effects of the covid-19 pandemic and related public health measures.
\newblock {\em The Lancet Regional Health-Americas}, 7:100130.

\bibitem[Every-Palmer et~al., 2020]{every2020psychological}
Every-Palmer, S., Jenkins, M., Gendall, P., Hoek, J., Beaglehole, B., Bell, C.,
  Williman, J., Rapsey, C., and Stanley, J. (2020).
\newblock Psychological distress, anxiety, family violence, suicidality, and
  wellbeing in new zealand during the covid-19 lockdown: A cross-sectional
  study.
\newblock {\em PLoS one}, 15(11):e0241658.

\bibitem[Fares, 2013]{fares2013winter}
Fares, A. (2013).
\newblock Winter cardiovascular diseases phenomenon.
\newblock {\em North American journal of medical sciences}, 5(4):266.

\bibitem[Frome, 1983]{frome1983analysis}
Frome, E.~L. (1983).
\newblock The analysis of rates using poisson regression models.
\newblock {\em Biometrics}, pages 665--674.

\bibitem[Gill and DeJoseph, 2020]{gill2020importance}
Gill, J.~R. and DeJoseph, M.~E. (2020).
\newblock The importance of proper death certification during the covid-19
  pandemic.
\newblock {\em Jama}, 324(1):27--28.

\bibitem[Gormsen and Koijen, 2020]{gormsen2020coronavirus}
Gormsen, N.~J. and Koijen, R.~S. (2020).
\newblock Coronavirus: Impact on stock prices and growth expectations.
\newblock {\em The Review of Asset Pricing Studies}, 10(4):574--597.

\bibitem[Gundlapalli et~al., 2021]{gundlapalli2021death}
Gundlapalli, A.~V., Lavery, A.~M., Boehmer, T.~K., Beach, M.~J., Walke, H.~T.,
  Sutton, P.~D., and Anderson, R.~N. (2021).
\newblock Death certificate--based icd-10 diagnosis codes for covid-19
  mortality surveillance—united states, january--december 2020.
\newblock {\em Morbidity and Mortality Weekly Report}, 70(14):523.

\bibitem[Hale et~al., 2021]{hale2021global}
Hale, T., Angrist, N., Goldszmidt, R., Kira, B., Petherick, A., Phillips, T.,
  Webster, S., Cameron-Blake, E., Hallas, L., Majumdar, S., et~al. (2021).
\newblock A global panel database of pandemic policies (oxford covid-19
  government response tracker).
\newblock {\em Nature Human Behaviour}, 5(4):529--538.

\bibitem[Hao et~al., 2020]{hao2020psychiatric}
Hao, F., Tan, W., Jiang, L., Zhang, L., Zhao, X., Zou, Y., Hu, Y., Luo, X.,
  Jiang, X., McIntyre, R.~S., et~al. (2020).
\newblock Do psychiatric patients experience more psychiatric symptoms during
  covid-19 pandemic and lockdown? a case-control study with service and
  research implications for immunopsychiatry.
\newblock {\em Brain, Behavior, and Immunity}, 87:100--106.

\bibitem[Hoff, 2015]{hoff2015multilinear}
Hoff, P.~D. (2015).
\newblock Multilinear tensor regression for longitudinal relational data.
\newblock {\em The annals of applied statistics}, 9(3):1169.

\bibitem[Hsieh et~al., 2020]{hsieh2020outcome}
Hsieh, C.-C., Lin, C.-H., Wang, W. Y.~C., Pauleen, D.~J., and Chen, J.~V.
  (2020).
\newblock The outcome and implications of public precautionary measures in
  taiwan--declining respiratory disease cases in the covid-19 pandemic.
\newblock {\em International Journal of Environmental Research and Public
  Health}, 17(13):4877.

\bibitem[Huh et~al., 2021]{huh2021decrease}
Huh, K., Kim, Y.-E., Ji, W., Kim, D.~W., Lee, E.-J., Kim, J.-H., Kang, J.-M.,
  and Jung, J. (2021).
\newblock Decrease in hospital admissions for respiratory diseases during the
  covid-19 pandemic: a nationwide claims study.
\newblock {\em Thorax}, 76(9):939--941.

\bibitem[Karlinsky and Kobak, 2021]{karlinsky2021tracking}
Karlinsky, A. and Kobak, D. (2021).
\newblock Tracking excess mortality across countries during the covid-19
  pandemic with the world mortality dataset.
\newblock {\em Elife}, 10:e69336.

\bibitem[Kepp et~al., 2022]{kepp2022estimates}
Kepp, K.~P., Bj{\"o}rk, J., Kontis, V., Parks, R.~M., B{\ae}k, K.~T., Emilsson,
  L., and Lallukka, T. (2022).
\newblock Estimates of excess mortality for the five nordic countries during
  the covid-19 pandemic 2020- 2021.
\newblock {\em International Journal of Epidemiology}, 51(6):1722--1732.

\bibitem[Kretzschmar et~al., 2020]{kretzschmar2020impact}
Kretzschmar, M.~E., Rozhnova, G., Bootsma, M.~C., van Boven, M., van~de
  Wijgert, J.~H., and Bonten, M.~J. (2020).
\newblock Impact of delays on effectiveness of contact tracing strategies for
  covid-19: a modelling study.
\newblock {\em The Lancet Public Health}, 5(8):e452--e459.

\bibitem[Kuzemko et~al., 2020]{kuzemko2020covid}
Kuzemko, C., Bradshaw, M., Bridge, G., Goldthau, A., Jewell, J., Overland, I.,
  Scholten, D., Van~de Graaf, T., and Westphal, K. (2020).
\newblock Covid-19 and the politics of sustainable energy transitions.
\newblock {\em Energy Research \& Social Science}, 68:101685.

\bibitem[Lippi and Henry, 2020]{lippi2020chronic}
Lippi, G. and Henry, B.~M. (2020).
\newblock Chronic obstructive pulmonary disease is associated with severe
  coronavirus disease 2019 (covid-19).
\newblock {\em Respiratory Medicine}, 167:105941.

\bibitem[Mehrizi et~al., 2021]{mehrizi2021trend}
Mehrizi, S., Vu, T.~X., Chatzinotas, S., and Ottersten, B. (2021).
\newblock Trend-aware proactive caching via tensor train decomposition: A
  bayesian viewpoint.
\newblock {\em IEEE Open Journal of the Communications Society}, 2:975--989.

\bibitem[Mitchell and Li, 2021]{mitchell2021state}
Mitchell, T.~O. and Li, L. (2021).
\newblock State-level data on suicide mortality during covid-19 quarantine:
  early evidence of a disproportionate impact on racial minorities.
\newblock {\em Psychiatry Research}, 295:113629.

\bibitem[Msemburi et~al., 2023]{msemburi2023estimates}
Msemburi, W., Karlinsky, A., Knutson, V., Aleshin-Guendel, S., Chatterji, S.,
  and Wakefield, J. (2023).
\newblock The who estimates of excess mortality associated with the covid-19
  pandemic.
\newblock {\em Nature}, 613(7942):130--137.

\bibitem[Oseledets, 2011]{oseledets2011tensor}
Oseledets, I.~V. (2011).
\newblock Tensor-train decomposition.
\newblock {\em SIAM Journal on Scientific Computing}, 33(5):2295--2317.

\bibitem[Pell et~al., 2020]{pell2020coronial}
Pell, R., Fryer, E., Manek, S., Winter, L., and Roberts, I.~S. (2020).
\newblock Coronial autopsies identify the indirect effects of covid-19.
\newblock {\em The Lancet Public Health}, 5(9):e474.

\bibitem[Peretti-Watel et~al., 2020]{peretti2020future}
Peretti-Watel, P., Seror, V., Cortaredona, S., Launay, O., Raude, J., Verger,
  P., Fressard, L., Beck, F., Legleye, S., l'Haridon, O., et~al. (2020).
\newblock A future vaccination campaign against covid-19 at risk of vaccine
  hesitancy and politicisation.
\newblock {\em The Lancet infectious diseases}, 20(7):769--770.

\bibitem[Phillips et~al., 2010]{phillips2010christmas}
Phillips, D., Barker, G.~E., and Brewer, K.~M. (2010).
\newblock Christmas and new year as risk factors for death.
\newblock {\em Social science \& medicine}, 71(8):1463--1471.

\bibitem[Pieh et~al., 2021]{pieh2021mental}
Pieh, C., Budimir, S., Delgadillo, J., Barkham, M., Fontaine, J.~R., and
  Probst, T. (2021).
\newblock Mental health during covid-19 lockdown in the united kingdom.
\newblock {\em Psychosomatic Medicine}, 83(4):328--337.

\bibitem[Roberts and Rosenthal, 2009]{roberts2009examples}
Roberts, G.~O. and Rosenthal, J.~S. (2009).
\newblock Examples of adaptive mcmc.
\newblock {\em Journal of Computational and Graphical Statistics},
  18(2):349--367.

\bibitem[Rossi et~al., 2020]{rossi2020covid}
Rossi, R., Socci, V., Talevi, D., Mensi, S., Niolu, C., Pacitti, F., Di~Marco,
  A., Rossi, A., Siracusano, A., and Di~Lorenzo, G. (2020).
\newblock Covid-19 pandemic and lockdown measures impact on mental health among
  the general population in italy.
\newblock {\em Frontiers in Psychiatry}, page 790.

\bibitem[Sarkodie and Owusu, 2021]{sarkodie2021global}
Sarkodie, S.~A. and Owusu, P.~A. (2021).
\newblock Global assessment of environment, health and economic impact of the
  novel coronavirus (covid-19).
\newblock {\em Environment, development and sustainability}, 23(4):5005--5015.

\bibitem[Schein et~al., 2015]{schein2015bayesian}
Schein, A., Paisley, J., Blei, D.~M., and Wallach, H. (2015).
\newblock Bayesian poisson tensor factorization for inferring multilateral
  relations from sparse dyadic event counts.
\newblock In {\em Proceedings of the 21th ACM SIGKDD International Conference
  on Knowledge Discovery and Data Mining}, pages 1045--1054.

\bibitem[Schein et~al., 2016]{schein2016bayesian}
Schein, A., Zhou, M., Blei, D., and Wallach, H. (2016).
\newblock Bayesian poisson tucker decomposition for learning the structure of
  international relations.
\newblock In {\em International Conference on Machine Learning}, pages
  2810--2819. PMLR.

\bibitem[Schwartz, 2000]{schwartz2000harvesting}
Schwartz, J. (2000).
\newblock Harvesting and long term exposure effects in the relation between air
  pollution and mortality.
\newblock {\em American Journal of Epidemiology}, 151(5):440--448.

\bibitem[Shiels et~al., 2022]{shiels2022leading}
Shiels, M.~S., Haque, A.~T., de~Gonz{\'a}lez, A.~B., and Freedman, N.~D.
  (2022).
\newblock Leading causes of death in the us during the covid-19 pandemic, march
  2020 to october 2021.
\newblock {\em JAMA Internal Medicine}, 182(8):883--886.

\bibitem[Stewart et~al., 2017]{stewart2017seasonal}
Stewart, S., Keates, A.~K., Redfern, A., and McMurray, J.~J. (2017).
\newblock Seasonal variations in cardiovascular disease.
\newblock {\em Nature Reviews Cardiology}, 14(11):654--664.

\bibitem[Sutherland et~al., 2020]{sutherland2020vehicle}
Sutherland, M., McKenney, M., and Elkbuli, A. (2020).
\newblock Vehicle related injury patterns during the covid-19 pandemic: what
  has changed?
\newblock {\em The American journal of emergency medicine}, 38(9):1710--1714.

\bibitem[Wang et~al., 2022]{wang2022estimating}
Wang, H., Paulson, K.~R., Pease, S.~A., Watson, S., Comfort, H., Zheng, P.,
  Aravkin, A.~Y., Bisignano, C., Barber, R.~M., Alam, T., et~al. (2022).
\newblock Estimating excess mortality due to the covid-19 pandemic: a
  systematic analysis of covid-19-related mortality, 2020--21.
\newblock {\em The Lancet}, 399(10334):1513--1536.

\bibitem[Zhao et~al., 2019]{zhao2019assessment}
Zhao, Q., Li, S., Coelho, M.~S., Saldiva, P.~H., Hu, K., Abramson, M.~J.,
  Huxley, R.~R., and Guo, Y. (2019).
\newblock Assessment of intraseasonal variation in hospitalization associated
  with heat exposure in brazil.
\newblock {\em JAMA Network Open}, 2(2):e187901--e187901.

\bibitem[Zhou et~al., 2015]{zhou2015bayesian}
Zhou, J., Bhattacharya, A., Herring, A.~H., and Dunson, D.~B. (2015).
\newblock Bayesian factorizations of big sparse tensors.
\newblock {\em Journal of the American Statistical Association},
  110(512):1562--1576.

\end{thebibliography}

\newpage

\appendix
\section*{Appendices}
\label{Appendices}

\section{Simulation study where parameters are artificial generated}
\label{simulation study I}

In the first experiment, we simulate count data from Poisson distribution with rate parameters generated according to the BPRTTD model. In this step, we fix $N=20,T=20$ and $K=20$ and the Tensor Train Decomposition rank $H_1=H_2=5$. We include one intercept plus $P=5$ covariates whose regression coefficients $\boldsymbol{\beta}$ are sampled from a normal distribution with mean 0 and variance 0.1. $\lambda^{(1)}_{i,h_1}$ are generated from a Gamma distribution with $a_\alpha=1$ and $b_\alpha=2.8$. $\lambda^{(2)}_{t,h_1,h_2}$ and $\lambda^{(3)}_{k,h_2}$ are simulated from the same Gamma distribution as well. Then with fixed parameter values, we generate covariates $\mathbf{x}_{i,t,k}$ from a standard normal distribution, offset $u_{i,t,k}$ from a Gamma distribution with shape and rate equal to 5 and 1. We repeat the simulation for 100 times. In each repetition, the observed data are simulated according to \eqref{eq:BPRTTD}. Finally, we apply the BPRTTD model to the simulated $Y_{i,t,k}$. In this step, we assume that true latent dimension $H_1$ and $H_2$ are known and we set up the parameters of the prior distribution according to the following values, $\alpha_a=1, \alpha_b=1, \beta_a=1, \beta_b=2, \epsilon_a=1, \epsilon_b=1$. The prior variance of $\boldsymbol{\beta}$ is 0.1. The probability of proposing from in the proposal distribution of the adaptive Metropolis-Hastings algorithm to update regression coefficients $p=0.05$. We run the MCMC for 10,000 iterations, discard the first 3,000 iteration. Results of comparison between true parameter values and the estimated ones are shown in Table \ref{tb:beta} and \Crefrange{tb:lambda1}{tb:lambda3}.

\begin{table}[ht]
\centering
\begin{tabular}{ccccccc}
\hline
$\boldsymbol{\beta}$ & -0.0387 & 0.1747 & 0.1103 & 0.1137 & 0.2840 & -0.6080 \\ \hline
$\hat{\boldsymbol{\beta}}$ & \begin{tabular}[c]{@{}c@{}}0.0078\\ (0.0342)\end{tabular} & \begin{tabular}[c]{@{}c@{}}0.1757\\ (0.0063)\end{tabular} & \begin{tabular}[c]{@{}c@{}}0.1097\\ (0.0057)\end{tabular} & \begin{tabular}[c]{@{}c@{}}0.1151\\ (0.0051)\end{tabular} & \begin{tabular}[c]{@{}c@{}}0.2848\\ (0.0067)\end{tabular} & \begin{tabular}[c]{@{}c@{}}-0.6093\\ (0.0054)\end{tabular} \\ \hline
\end{tabular}
\caption{Comparison between true $\boldsymbol{\beta}$ and the estimated $\hat{\boldsymbol{\beta}}$ from the BPRTTD model in terms of posterior mean. Averages and standard deviations (in parentheses) of the posterior means over 100 repetitions are reported.}
\label{tb:beta}
\end{table}

In \cref{tb:beta}, coefficients associated with simulated covariates $\mathbf{x}_{i,t,k}$ are estimated accurately with small standard deviations over 100 repetition of simulation studies. The estimated intercept has a higher standard deviation. This is due to the identifiability issue associated with the intercept and $\lambda^{(1)}_{i,h_1}, \lambda^{(2)}_{t,h_1,h_2}, \lambda^{(3)}_{k,h_2}$ inherent to the BPRTTD model as these parameters multiply and contribute to the Poisson rate. Careful choice of prior parameters helps overcome the identifiability problem and facilitate our goal to interpret factors $\lambda^{(1)}_{i,h_1}, \lambda^{(2)}_{t,h_1,h_2}, \lambda^{(3)}_{k,h_2}$. In fact, \Crefrange{tb:lambda1}{tb:lambda3} show how these parameters are recovered using our method. For $\lambda^{(1)}_{i,h_1}$, the difference between the 100 true values and their estimates by posterior means has mean 0.0148. This number is 0.0562 and -0.0675 for $\lambda^{(2)}_{t,h_1,h_2}$ and $\lambda^{(3)}_{k,h_2}$ respectively, validating our approach's ability to recover parameters for further analysis.

\begin{table}[ht]
\centering
\resizebox{\textwidth}{!}{

}
\caption{Comparison between true $\lambda^{(1)}_{i,h_1}, i=1,\dots,N, h_1=1,\dots,H_1$ and the estimated values from the BPRTTD model in terms of posterior mean. Average (\textit{italic}) and standard deviation (\textit{italic} in parentheses) of the posterior means over 100 repetitions are reported.}
\label{tb:lambda1}
\end{table}

\begin{table}[ht]
\centering
\resizebox{\textwidth}{!}{
%
}
\caption{Comparison between true $\lambda^{(2)}_{t,h_1,h_2}, t=1,\dots,T, h_1=1,\dots,H_1, h_2=1,\dots,H_2$ and the estimated values from the BPRTTD model in terms of posterior mean. Average (\textit{italic}) and standard deviation (\textit{italic} in parentheses) of the posterior means over 100 repetitions are reported.}
\label{tb:lambda2}
\end{table}

\begin{table}[ht]
\centering
\resizebox{\textwidth}{!}{
%
 \\ 
   \bottomrule
\end{tabular}
}
\caption{Comparison between true $\lambda^{(3)}_{k,h_2}, k=1,\dots,K, h_2=1,\dots,H_2$ and the estimated values from the BPRTTD model in terms of posterior mean. Average (\textit{italic}) and standard deviation (\textit{italic} in parentheses) of the posterior means over 100 repetitions are reported.}
\label{tb:lambda3}
\end{table}

\clearpage

\section{Poisson regression coefficients}
\label{Poisson regression coefficients}

The following table displays the posterior mean estimates of the Poisson regression coefficients in the BPRTTD model as well as the 95\% credible intervals.

\begin{sidewaystable}
    \centering
\resizebox{\textwidth}{!}{
\begin{tabular}{lccc|lccc|lccc|lccc}
  \hline
\multicolumn{1}{c}{Coefficients} & Posterior mean & 95\% CI lower bound & 95\% CI upper bound & \multicolumn{1}{c}{Coefficients} & Posterior mean & 95\% CI lower bound & 95\% CI upper bound & \multicolumn{1}{c}{Coefficients} & Posterior mean & 95\% CI lower bound & 95\% CI upper bound & \multicolumn{1}{c}{Coefficients} & Posterior mean & 95\% CI lower bound & 95\% CI upper bound \\ 
  \hline
Intercept & -16.6495 & -16.6527 & -16.6462 & causes8:age50-59 & 0.8084 & 0.8063 & 0.8110 & causes9:age70-74 & 1.9028 & 1.9003 & 1.9054 & causes10:age85-89 & 1.9912 & 1.9871 & 1.9961 \\ 
  isi & 0.0622 & 0.0622 & 0.0623 & causes9:age50-59 & 0.6622 & 0.6597 & 0.6653 & causes10:age70-74 & 1.0739 & 1.0697 & 1.0788 & causes11:age85-89 & 3.5340 & 3.5208 & 3.5482 \\ 
  causes2 & -0.6237 & -0.6263 & -0.6212 & causes10:age50-59 & 0.8108 & 0.8066 & 0.8155 & causes11:age70-74 & 1.8440 & 1.8273 & 1.8612 & causes12:age85-89 & 2.2547 & 2.2510 & 2.2590 \\ 
  causes3 & 1.8678 & 1.8639 & 1.8717 & causes11:age50-59 & 0.9046 & 0.8903 & 0.9187 & causes12:age70-74 & 1.1411 & 1.1357 & 1.1473 & causes13:age85-89 & 3.8594 & 3.8573 & 3.8618 \\ 
  causes4 & -2.0146 & -2.0214 & -2.0086 & causes12:age50-59 & 0.5281 & 0.5240 & 0.5331 & causes13:age70-74 & 2.0869 & 2.0844 & 2.0897 & causes14:age85-89 & -3.1789 & -4.1169 & -2.3403 \\ 
  causes5 & -0.5788 & -0.5821 & -0.5758 & causes13:age50-59 & 0.9430 & 0.9406 & 0.9456 & causes14:age70-74 & -5.2264 & -6.4161 & -4.2635 & causes15:age85-89 & -6.4308 & -6.4654 & -6.3910 \\ 
  causes6 & -1.4547 & -1.4584 & -1.4512 & causes14:age50-59 & -3.1612 & -3.1970 & -3.1241 & causes15:age70-74 & -6.3054 & -6.3257 & -6.2844 & causes16:age85-89 & -2.1175 & -2.1227 & -2.1124 \\
\hline
\end{tabular}
}
\caption{Posterior mean estimates $\hat{\boldsymbol{\beta}}$ of the BPRTTD model with 95\% credible intervals.}
\label{tb:coefficients}
\end{sidewaystable}

\clearpage

\section{Latent parameters}
\label{Latent parameters}

The following table displays the posterior mean estimates of the tensor train decomposition parameters $\lambda^{(1)}_{i,h_1}$ in the BPRTTD model. 

\begin{table}[ht]\captionsetup[subtable]{font=tiny}
\centering
\begin{subtable}[t]{0.6\textheight}
\resizebox{\textwidth}{!}{
\begin{tabular}{lllllllllllllllllllll}
  \hline
  & Male 00-49 & Male 50-59 & Male 60-64 & Male 65-69 & Male 70-74 & Male 75-79 & Male 80-84 & Male 85-89 & Male 90-94 & Male 95+ & Female 00-49 & Female 50-59 & Female 60-64 & Female 65-69 & Female 70-74 & Female 75-79 & Female 80-84 & Female 85-89 & Female 90-94 & Female 95+ \\ 
  \hline
Piemonte & \color{red}0.2176 & \color{red}0.1779 & \color{red}0.2456 & \color{black}0.1583 & \color{black}0.1277 & \color{black}0.1419 & \color{black}0.1351 & \color{black}0.1061 & \color{black}0.0799 & \color{black}0.0431 & \color{black}0.1312 & \color{red}0.2436 & \color{red}0.2442 & \color{red}0.4634 & \color{red}0.4536 & \color{red}0.4853 & \color{red}0.4126 & \color{red}0.3386 & \color{red}0.2555 & \color{red}0.2102 \\ 
  Valle d'Aosta & \color{black}0.0926 & \color{black}0.1480 & \color{black}0.1575 & \color{red}0.2142 & \color{red}0.1906 & \color{red}0.1744 & \color{red}0.3431 & \color{red}0.3038 & \color{red}0.1781 & \color{black}0.1307 & \color{black}0.1568 & \color{black}0.1502 & \color{red}0.1883 & \color{red}0.1688 & \color{red}0.2560 & \color{red}0.3198 & \color{red}0.4968 & \color{red}0.4399 & \color{red}0.5008 & \color{red}0.4501 \\ 
  Lombardia & \color{black}0.1580 & \color{red}0.1784 & \color{black}0.0874 & \color{black}0.0688 & \color{black}0.0531 & \color{black}0.0427 & \color{black}0.0786 & \color{black}0.0836 & \color{black}0.0991 & \color{black}0.0644 & \color{black}0.0989 & \color{black}0.1233 & \color{red}0.2349 & \color{red}0.3042 & \color{red}0.3308 & \color{red}0.3454 & \color{red}0.3565 & \color{red}0.3446 & \color{red}0.3095 & \color{red}0.3090 \\
\hline
\end{tabular}
}
\caption{$\hat{\lambda}^{(1)}_{i,1}$}
\label{tb:lambda1_real1}
\end{subtable}
\begin{subtable}[t]{0.6\textheight}
\resizebox{\textwidth}{!}{
\begin{tabular}{lllllllllllllllllllll}
  \hline
  & Male 00-49 & Male 50-59 & Male 60-64 & Male 65-69 & Male 70-74 & Male 75-79 & Male 80-84 & Male 85-89 & Male 90-94 & Male 95+ & Female 00-49 & Female 50-59 & Female 60-64 & Female 65-69 & Female 70-74 & Female 75-79 & Female 80-84 & Female 85-89 & Female 90-94 & Female 95+ \\ 
  \hline
Piemonte & \color{black}0.0089 & \color{black}0.0109 & \color{black}0.0238 & \color{black}0.0316 & \color{black}0.0294 & \color{black}0.0362 & \color{black}0.0907 & \color{black}0.1551 & \color{red}0.2000 & \color{red}0.2607 & \color{black}0.0255 & \color{black}0.0169 & \color{black}0.0426 & \color{black}0.0509 & \color{black}0.1417 & \color{red}0.1882 & \color{red}0.2719 & \color{red}0.3593 & \color{red}0.3935 & \color{red}0.3755 \\ 
  Valle d'Aosta & \color{black}0.1243 & \color{black}0.1338 & \color{black}0.1148 & \color{black}0.1229 & \color{black}0.1238 & \color{black}0.0998 & \color{black}0.0977 & \color{black}0.1198 & \color{black}0.1042 & \color{black}0.1258 & \color{black}0.1501 & \color{black}0.0978 & \color{black}0.1301 & \color{black}0.1147 & \color{black}0.1123 & \color{black}0.1395 & \color{black}0.1525 & \color{black}0.0993 & \color{black}0.1654 & \color{black}0.1614 \\ 
  Lombardia & \color{black}0.0078 & \color{black}0.0167 & \color{black}0.0171 & \color{black}0.0164 & \color{black}0.0122 & \color{black}0.0140 & \color{black}0.0185 & \color{black}0.0361 & \color{black}0.0572 & \color{black}0.0501 & \color{black}0.0112 & \color{black}0.0090 & \color{black}0.0224 & \color{black}0.0140 & \color{black}0.0398 & \color{black}0.0739 & \color{black}0.1339 & \color{black}0.1517 & \color{black}0.1470 & \color{black}0.1315 \\
\hline
\end{tabular}
}
\caption{$\hat{\lambda}^{(1)}_{i,2}$}
\label{tb:lambda1_real2}
\end{subtable}
\begin{subtable}[t]{0.6\textheight}
\resizebox{\textwidth}{!}{
\begin{tabular}{lllllllllllllllllllll}
  \hline
  & Male 00-49 & Male 50-59 & Male 60-64 & Male 65-69 & Male 70-74 & Male 75-79 & Male 80-84 & Male 85-89 & Male 90-94 & Male 95+ & Female 00-49 & Female 50-59 & Female 60-64 & Female 65-69 & Female 70-74 & Female 75-79 & Female 80-84 & Female 85-89 & Female 90-94 & Female 95+ \\ 
  \hline
Piemonte & \color{black}0.0531 & \color{black}0.1042 & \color{black}0.1188 & \color{black}0.0778 & \color{black}0.1016 & \color{black}0.1337 & \color{black}0.1417 & \color{black}0.1140 & \color{black}0.0777 & \color{black}0.0238 & \color{black}0.0812 & \color{black}0.0251 & \color{black}0.0347 & \color{black}0.0462 & \color{black}0.0550 & \color{black}0.0696 & \color{black}0.1190 & \color{red}0.2297 & \color{red}0.2887 & \color{red}0.3435 \\ 
  Valle d'Aosta & \color{black}0.1128 & \color{black}0.0972 & \color{black}0.1257 & \color{black}0.1640 & \color{black}0.1139 & \color{black}0.1328 & \color{red}0.1712 & \color{black}0.1546 & \color{black}0.0859 & \color{black}0.1105 & \color{black}0.1363 & \color{black}0.1147 & \color{black}0.1237 & \color{black}0.0812 & \color{black}0.1463 & \color{black}0.1624 & \color{black}0.1544 & \color{red}0.2364 & \color{red}0.2585 & \color{red}0.3347 \\ 
  Lombardia & \color{black}0.0681 & \color{black}0.0398 & \color{black}0.0284 & \color{black}0.0301 & \color{black}0.0445 & \color{black}0.0485 & \color{black}0.0635 & \color{black}0.0297 & \color{black}0.0060 & \color{black}0.0047 & \color{black}0.0415 & \color{black}0.0254 & \color{black}0.0152 & \color{black}0.0138 & \color{black}0.0106 & \color{black}0.0108 & \color{black}0.0232 & \color{black}0.0583 & \color{black}0.0900 & \color{black}0.1469 \\
\hline
\end{tabular}
}
\caption{$\hat{\lambda}^{(1)}_{i,3}$}
\label{tb:lambda1_real3}
\end{subtable}
\begin{subtable}[t]{0.6\textheight}
\resizebox{\textwidth}{!}{
\begin{tabular}{lllllllllllllllllllll}
  \hline
  & Male 00-49 & Male 50-59 & Male 60-64 & Male 65-69 & Male 70-74 & Male 75-79 & Male 80-84 & Male 85-89 & Male 90-94 & Male 95+ & Female 00-49 & Female 50-59 & Female 60-64 & Female 65-69 & Female 70-74 & Female 75-79 & Female 80-84 & Female 85-89 & Female 90-94 & Female 95+ \\ 
  \hline
Piemonte & \color{red}0.4170 & \color{red}0.4056 & \color{red}0.3023 & \color{red}0.2667 & \color{red}0.2324 & \color{red}0.1878 & \color{black}0.1423 & \color{black}0.1172 & \color{black}0.1091 & \color{black}0.1194 & \color{black}0.0281 & \color{black}0.0316 & \color{black}0.0705 & \color{black}0.0921 & \color{black}0.0543 & \color{black}0.0530 & \color{black}0.0482 & \color{black}0.0137 & \color{black}0.0155 & \color{black}0.0271 \\ 
  Valle d'Aosta & \color{red}0.6247 & \color{red}0.4970 & \color{red}0.2679 & \color{red}0.3391 & \color{red}0.2693 & \color{black}0.1550 & \color{black}0.1538 & \color{black}0.0919 & \color{black}0.0781 & \color{black}0.1594 & \color{black}0.1438 & \color{black}0.1072 & \color{black}0.1294 & \color{black}0.1307 & \color{black}0.1547 & \color{black}0.1373 & \color{black}0.0888 & \color{black}0.0841 & \color{black}0.0404 & \color{black}0.0572 \\ 
  Lombardia & \color{red}0.3005 & \color{red}0.2369 & \color{red}0.1946 & \color{black}0.1430 & \color{black}0.1119 & \color{black}0.0946 & \color{black}0.0782 & \color{black}0.0908 & \color{black}0.0925 & \color{black}0.1210 & \color{black}0.0093 & \color{black}0.0096 & \color{black}0.0167 & \color{black}0.0135 & \color{black}0.0188 & \color{black}0.0170 & \color{black}0.0138 & \color{black}0.0230 & \color{black}0.0320 & \color{black}0.0380 \\
\hline
\end{tabular}
}
\caption{$\hat{\lambda}^{(1)}_{i,4}$}
\label{tb:lambda1_real4}
\end{subtable}
\begin{subtable}[t]{0.6\textheight}
\resizebox{\textwidth}{!}{
\begin{tabular}{lllllllllllllllllllll}
  \hline
  & Male 00-49 & Male 50-59 & Male 60-64 & Male 65-69 & Male 70-74 & Male 75-79 & Male 80-84 & Male 85-89 & Male 90-94 & Male 95+ & Female 00-49 & Female 50-59 & Female 60-64 & Female 65-69 & Female 70-74 & Female 75-79 & Female 80-84 & Female 85-89 & Female 90-94 & Female 95+ \\ 
  \hline
Piemonte & \color{black}0.1311 & \color{red}0.2530 & \color{red}0.2568 & \color{red}0.3332 & \color{red}0.3715 & \color{red}0.3146 & \color{red}0.2959 & \color{red}0.2664 & \color{red}0.1741 & \color{black}0.1136 & \color{black}0.0972 & \color{black}0.1082 & \color{red}0.2215 & \color{black}0.1205 & \color{red}0.2086 & \color{red}0.2733 & \color{red}0.2495 & \color{red}0.2337 & \color{red}0.2347 & \color{red}0.1688 \\ 
  Valle d'Aosta & \color{black}0.0960 & \color{black}0.1146 & \color{red}0.2494 & \color{red}0.1752 & \color{red}0.2103 & \color{red}0.3083 & \color{black}0.1597 & \color{black}0.1022 & \color{black}0.1653 & \color{black}0.1520 & \color{black}0.1459 & \color{red}0.1968 & \color{black}0.1468 & \color{black}0.1428 & \color{black}0.1520 & \color{red}0.1873 & \color{black}0.1397 & \color{red}0.2151 & \color{black}0.1335 & \color{black}0.1025 \\ 
  Lombardia & \color{red}0.3268 & \color{red}0.5206 & \color{red}0.6397 & \color{red}0.7222 & \color{red}0.6937 & \color{red}0.6661 & \color{red}0.5534 & \color{red}0.4270 & \color{red}0.2946 & \color{black}0.1565 & \color{red}0.2078 & \color{red}0.2384 & \color{red}0.2701 & \color{red}0.3549 & \color{red}0.4463 & \color{red}0.5078 & \color{red}0.4943 & \color{red}0.4982 & \color{red}0.5259 & \color{red}0.4702 \\
\hline
\end{tabular}
}
\caption{$\hat{\lambda}^{(1)}_{i,5}$}
\label{tb:lambda1_real5}
\end{subtable}
\begin{subtable}[t]{0.6\textheight}
\resizebox{\textwidth}{!}{
\begin{tabular}{lllllllllllllllllllll}
  \hline
  & Male 00-49 & Male 50-59 & Male 60-64 & Male 65-69 & Male 70-74 & Male 75-79 & Male 80-84 & Male 85-89 & Male 90-94 & Male 95+ & Female 00-49 & Female 50-59 & Female 60-64 & Female 65-69 & Female 70-74 & Female 75-79 & Female 80-84 & Female 85-89 & Female 90-94 & Female 95+ \\ 
  \hline
Piemonte & \color{black}0.0678 & \color{black}0.0889 & \color{black}0.0979 & \color{black}0.1430 & \color{black}0.1657 & \color{red}0.2039 & \color{red}0.2426 & \color{red}0.3120 & \color{red}0.4211 & \color{red}0.4941 & \color{red}0.5065 & \color{red}0.3893 & \color{red}0.3005 & \color{red}0.1815 & \color{black}0.1281 & \color{black}0.0518 & \color{black}0.0421 & \color{black}0.0142 & \color{black}0.0147 & \color{black}0.0217 \\ 
  Valle d'Aosta & \color{black}0.0496 & \color{black}0.0646 & \color{black}0.1495 & \color{black}0.0979 & \color{red}0.1680 & \color{red}0.1942 & \color{red}0.1943 & \color{red}0.2789 & \color{red}0.4678 & \color{red}0.2977 & \color{red}0.1994 & \color{red}0.1844 & \color{red}0.2606 & \color{red}0.2209 & \color{red}0.1952 & \color{black}0.1124 & \color{black}0.0632 & \color{black}0.0381 & \color{black}0.0332 & \color{black}0.0763 \\ 
  Lombardia & \color{black}0.0529 & \color{black}0.0374 & \color{black}0.0969 & \color{black}0.1163 & \color{black}0.1603 & \color{red}0.2093 & \color{red}0.2708 & \color{red}0.3730 & \color{red}0.4790 & \color{red}0.6461 & \color{red}0.4626 & \color{red}0.3926 & \color{red}0.3005 & \color{red}0.2239 & \color{black}0.1554 & \color{black}0.1129 & \color{black}0.0735 & \color{black}0.0505 & \color{black}0.0526 & \color{black}0.0848 \\
\hline
\end{tabular}
}
\caption{$\hat{\lambda}^{(1)}_{i,6}$}
\label{tb:lambda1_real6}
\end{subtable}
\caption{Posterior mean estimates of $\lambda^{(1)}_{i,h_1}, i=1,\dots,N, h_1=1,\dots,H_1$ from the BPRTTD model. Red colored numbers indicate that the estimates are higher than the prior mean.}
\label{tb:lambda1_real}
\end{table}


\end{document}